\documentclass[aps,prb,reprint,superscriptaddress]{revtex4-1}
\usepackage{lmodern}
\usepackage{amssymb,amsmath}
\usepackage{ifxetex,ifluatex}
\usepackage{fixltx2e} 
\ifnum 0\ifxetex 1\fi\ifluatex 1\fi=0 
  \usepackage[T1]{fontenc}
  \usepackage[utf8]{inputenc}
\else 
  \ifxetex
    \usepackage{mathspec}
  \else
    \usepackage{fontspec}
  \fi
  \defaultfontfeatures{Ligatures=TeX,Scale=MatchLowercase}
\fi
\IfFileExists{upquote.sty}{\usepackage{upquote}}{}
\IfFileExists{microtype.sty}{%
\usepackage[]{microtype}
\UseMicrotypeSet[protrusion]{basicmath} 
}{}
\PassOptionsToPackage{hyphens}{url} 
\usepackage[unicode=true]{hyperref}
\PassOptionsToPackage{usenames,dvipsnames}{color} 
\hypersetup{
            pdftitle={High-Temperature Majorana Fermions in Magnet-Superconductor Hybrid Systems},
            colorlinks=true,
            linkcolor=blue,
            citecolor=blue,
            urlcolor=blue,
            breaklinks=true,
			final=true}
\usepackage{natbib}
\usepackage{graphicx,grffile}
\makeatletter
\def\maxwidth{\ifdim\Gin@nat@width>\linewidth\linewidth\else\Gin@nat@width\fi}
\def\maxheight{\ifdim\Gin@nat@height>\textheight\textheight\else\Gin@nat@height\fi}
\makeatother
\setkeys{Gin}{width=\maxwidth,height=\maxheight,keepaspectratio}
\ifx\paragraph\undefined\else
\let\oldparagraph\paragraph
\renewcommand{\paragraph}[1]{\oldparagraph{#1}\mbox{}}
\fi
\ifx\subparagraph\undefined\else
\let\oldsubparagraph\subparagraph
\renewcommand{\subparagraph}[1]{\oldsubparagraph{#1}\mbox{}}
\fi

\makeatletter
\def\fps@figure{htbp}
\makeatother

\usepackage{xparse}
\usepackage{xcolor}

\renewcommand{\vec}[1]{\mathbf{#1}}

\def\lparen{(}
\def\rparen{)}
\catcode`(=\active
\catcode`)=\active
\newcommand{(}{\ifmmode\left\lparen\else\lparen\fi}
\newcommand{)}{\ifmmode\right\rparen\else\rparen\fi}

\def\equationautorefname~#1\null{%
  Eq.~(#1)\null
}

\def\figureautorefname~#1\null{%
  Fig.~#1\null
}

\def\sectionautorefname~#1\null{%
	Sec.~#1\null
}

\newcommand{\up}{{\uparrow}}
\newcommand{\dw}{{\downarrow}}

\newcommand{\new}[1]{{\color{black}#1}}
\newcommand{\eps}{{\varepsilon}}
\newcommand{\nn}{\nonumber}
\newcommand{\ie}{{\it i.e.},\ }
\def\eg{\emph{e.g.}\ }
\def\ea{\emph{et al.}}

\begin{document}
\title{High-Temperature Majorana Fermions in Magnet-Superconductor Hybrid
Systems}

\author{Daniel Crawford}
\affiliation{School of Physics, University of Melbourne, Parkville, VIC 3010,
Australia}
\author{Eric Mascot}
\affiliation{University of Illinois at Chicago, Chicago, IL 60607, USA}
\author{Dirk K. Morr}
\affiliation{University of Illinois at Chicago, Chicago, IL 60607, USA}
\author{Stephan Rachel}
\affiliation{School of Physics, University of Melbourne, Parkville, VIC 3010,
Australia}

\begin{abstract}
Magnet-superconductor hybrid (MSH) structures represent one of the most
promising platforms to realize, control and manipulate Majorana modes
using scanning tunneling methods. By depositing either chains or islands
of magnetic atoms on the surface of a conventional, elemental superconductor such
as Pb or Re, topological superconducting phases can emerge. They feature
either localised Majorana bound states at the chain ends or dispersing
chiral Majorana modes at the island's boundary. Yet some of these
experiments have not reached the spectral resolution to clearly
	distinguish between topological Majorana and trivial Shiba states due to
very small superconducting gap sizes and experiments performed at sub-Kelvin temperatures.
Here we consider superconducting substrates with unconventional
spin-singlet pairing, including high-temperature
\(d\)-wave and extended \(s\)-wave superconductors. We derive
topological phase diagrams and compute edge states for cylinder and
island geometries and discuss their properties. Several
time-reversal invariant topological superconducting phases of the
	Zhang-Kane-Mele type are found and discussed. Moreover, we study one-dimensional 
MSH structures and show that parameters to realize topologically non-trivial magnetic chains embedded into a larger, two-dimensional substrate differ from the purely one-dimensional case.
Quite generally we find
that unconventional superconducting substrates work as well as the
conventional $s$-wave substrates to realize topological phases. In
particular, iron-based pnictide and chalcogenide superconductors are the
most promising class of substrates for future high-temperature
MSH systems.
\end{abstract}

\maketitle

\hypertarget{introduction}{%
\section{Introduction}\label{introduction}}

Majorana zero-modes in topological superconductors (TSCs) have attracted
a great deal of interest in recent years because of potential
applications in fault-tolerant quantum computing\,\citep{moore_nonabelions_1991, read_paired_2000, ivanov_non-abelian_2001, nayak_non-abelian_2008}.
Majorana zero-modes are theoretically predicted to be non-Abelian
anyons; it is this exotic exchange statistics which is responsible for
the vision to combine Majorana zero-modes into qubits for topological
quantum computing. Information processing is topologically protected and
stored nonlocally. These qubits could then resolve the problem of short
coherence times leading to \emph{fault tolerant} quantum computing.

A Majorana zero-mode can be seen as the condensed matter equivalent of
the Majorana fermion --- a fundamental particle originally proposed in
1937 by E. Majorana\,\citep{majorana_teoria_1937} as a real solution to the Dirac equation (and a Majorana fermion is thus identical to its anti-particle). Majorana modes have
been proposed to arise in the $\nu=5/2$ fractional quantum Hall state\,\citep{moore_nonabelions_1991}, in superfluid He-3\,\citep{volovik_an_1988}, in fractionalised spin liquids\,\citep{kitaev_anyons_2006}, in  topological superconductors\,\citep{kallin_chiral_2016,sato-17rpp076501} and in superconducting  heterostructures\,\citep{PhysRevLett.105.077001,mourik_signatures_2012, nadj-perge_observation_2014, ruby_end_2015, pawlak_probing_2016}.

%
%
Chiral superconductors --- in particular, the \(p_x+ip_y\)-wave superconductor
--- are topologically non-trivial and the prototypical candidate for
hosting Majorana zero-modes\,\citep{kitaev_unpaired_2001}. The chiral
$p$-wave pairing is rare in nature\,\citep{sato-17rpp076501};
Sr$_2$RuO$_4$ and UPt$_3$ are two well-established (and controversially discussed) candidates, 
although recent experiments on Sr$_2$RuO$_4$ seem to hint towards a \emph{helical} pairing
state instead\,\citep{kashiwaya-19prb094530}. More recent candidate materials include Cu$_x$Bi$_2$Se$_{3}$\,\cite{hor-10prl057001} and Sn$_{1-x}$In$_x$Te\,\cite{erickson-09prb024520,novak-13prb140502}.
Chiral superconductivity can be engineered using heterostructures involving \emph{proximity-induced} superconductivity. Recently, tremendous progress has been made to realise such systems. These structures have the major advantage that they are made out of
well-characterised and well-controlled ingredients, leading to a remarkable
experimental accessibility.

Proposals for engineering topological 
superconductors, all involving proximity-induced $s$-wave
superconductivity, include  superconductor--topological
insulator heterostructures\,\citep{fu_superconducting_2008}, coupled Rashba
nanowire--superconductors\, \citep{PhysRevLett.105.077001,mourik_signatures_2012, nadj-perge_observation_2014, ruby_end_2015, pawlak_probing_2016}, and ferromagnet-superconductor hybrid structures\,\citep{nadj-perge_observation_2014, li_topological_2014, li_two-dimensional_2016,kim_toward_2018,palacio-morales_atomic-scale_2019}.
Typically these proposals combine ingredients which spin-polarize the
electrons (such as magnetic moments or a magnetic field) and which mix
the spin multiplets (such as Rashba spin orbit coupling); together with the
proximity-induced superconductivity this leads to an effectively topological
phase. In the aforementioned ferromagnet-superconductor hybrid structures, ferromagnetic adatoms are
deposited on a superconductor as chains, lattices, or islands via self-assembly, epitaxial growth or single-atom manipulation techniques. In these
structures (Shiba chains, lattices, or islands, respectively) Yu-Shiba-Rusinov\,\cite{balatsky-06rmp373}
states arise at each ferromagnetic impurity. Adjacent Yu-Shiba-Rusinov states can hybridise and, if sufficiently many impurities are placed in vicinity, a band starts to form which can or cannot be topologically non-trivial. 
In 2D, the resulting topological superconductor is
characterised by a Chern number and chiral edge modes, as a
superconducting analogue of the integer quantum Hall effect (indeed, one might think of a chiral superconductor as a quantum Hall effect of superconducting Bogoliubov quasiparticles). The chiral
edge states at zero-energy constitute the one-dimensional, delocalized version of the desired Majorana zero-modes and complement the zero-dimensional Majorana bound states as present at vortex cores  or at the ends of one-dimensional topological superconductors.

Majorana zero modes (MZMs) were first reported in InSb nanowires at the interface to a superconductor\,\cite{mourik_signatures_2012}.
Majorana bound states at the end of Shiba chains were first reported in Fe/Pb(110)\,\cite{nadj-perge_observation_2014,ruby_end_2015,pawlak_probing_2016} and later in Fe/Re(0001)\,\cite{kim_toward_2018}. Recently, MZMs bound to vortex cores in the vortex lattice phase of FeSeTe have attracted considerable attention\,\cite{wang_evidence_2018, zhang_observation_2018}. Chiral Majorana modes were reported in Pb/Co/Si(111)\,\cite{menard_two-dimensional_2017} --- a somewhat special system since the magnetic Co atoms are sitting {\it below} the superconducting Pb layer \new{and it is thus difficult to identify the exact location of the Co island.} Fe islands on Re(0001)-O(2$\times$1) substrates represent a prototype of a Shiba island and chiral Majorana modes have been imaged\,\cite{palacio-morales_atomic-scale_2019}.
All the mentioned 
experiments involve scanning tunneling measurements in ultra-high vacuum
at sub-Kelvin temperatures. In a non-trivial phase edge states may be
detected by characteristic robust zero-bias conductance peaks, which
disappear when parameters are tuned to a trivial phase. In principle, atomic
manipulation techniques can be applied to manipulate and control the magnetic adatoms as well as their magnetic properties, leading to a variety of proposals\,\citep{mascot_quantum_2018,rachel_quantized_2017,mascot-19prb184510}.

The mentioned experiments are highly challenging and require very low
temperatures; additionally, the
spectral resolution is often not sufficient to unambiguously separate Majorana zero modes from
trivial Shiba states. Thus it would be most desirable to reproduce similar experiments on superconducting substrates with large gap sizes as realized in some of the high-temperature superconductors.
For example, the Re(0001)-O(2$\times$1) substrate has a gap size of 280---330
$\mu$eV\,\cite{palacio-morales_atomic-scale_2019} whereas superconducting gaps of up to 20 meV have been reported in monolayer FeSe\,\citep{ge_evidence_2019, jandke_unconventional_2019} and also up to 20
meV in LaFeAsO and LaFePO\,\citep{ishida_unusual_2008}. Obvious
candidates are the high \(T_c\) cuprate and iron-based superconductors as substrates for Shiba systems.

The cuprates are well known to have nodal lines in their superconducting
order parameter; we will consider, however, scenarios with fully gapped
\(d\)-wave superconductors. Nonetheless, the generic gaplessness of
cuprates motivates us to look for other  high-\(T_c\) systems such
as  iron pnictides and iron chalcogenides featuring extended \(s\)-wave pairing. 
Non-magnetic impurities on extended $s$-wave substrates were recently studied in Ref.\,\onlinecite{mashkoori-19prb014508}.
To complement these considerations, we also study substrates with chiral singlet pairings, \(d+id\) and \(s+id\).
These \(d\)-wave variants are often associated with water-intercalated
cobaltates \citep{takada_superconductivity_2003, PhysRevLett.91.097003}
but also 
twisted bilayer graphene \citep{dai_twisted_2016} and some iron pnictides\,\citep{chen_iron-based_2014, reid_d-wave_2012, kuroki_unconventional_2008, shen_observation_2019}. Highly
overdoped monolayer CuO\(_2\) constitutes another proposal for an
extended \(s\)-wave superconductor\,\citep{jiang_nodeless_2018}.

In this paper we investigate magnet-superconductor heterostructures
(MSH) with unconventional superconducting substrates featuring
spin-singlet pairing from a theory perspective. We introduce model and method in Sec.\,\ref{model-and-method}. We then briefly show well-established results for isotropic, nodeless $s$-wave substrates in Sec.\,\ref{s_000-s-wave} before we 
consider \(d\)-wave pairings as
realized in cuprate superconductors in Sec.\,\ref{d-wave} and test extensions and variations
of them in Sec.\,\ref{did}. Then we focus on various types and combinations of extended
\(s\)-wave pairings in Sec.\,\ref{extended-s-wave}. 
Due to the magnetic moments of the
impurities, time-reversal symmetry is explicitly broken and chiral
superconductivity expected to occur in these structures. In the absence of magnetic
impurities, we observe time-reversal invariant (TRI) TSC of the
Zhang-Kane-Mele type\,\cite{zhang_time-reversal-invariant_2013} in Sec.\,\ref{tri-phases}. 
In Sec.\,\ref{results1D}, we switch to one spatial dimension and investigate Shiba chains on a few selected unconventional substrates as well as their MZM end states. We compare the purely 1D case with the chains which are embedded into a 2D substrate in Sec.\,\ref{sec:chain-in-2D}, highly relevant to perform
topological quantum computing in network geometries\,\citep{alicea_non-abelian_2011}. 
We show that both systems behave differently and simulating purely 1D geometries fails to describe the experiments. In Sec.\,\ref{anti-shiba-chain} we introduce {\it anti-Shiba chains} as a system which realizes the 1D TRI topological phase in an experimentally relevant system.
\new{We show in Sec.\,\ref{topological-protection-of-tri-tsc-edge-states} and \ref{anti-shiba-chain} that the TRI TSC phase features robust MZMs in the presence of random disorder, even when time-reversal symmetry is broken explicitly.}
In Sec.\,\ref{discussion} we discuss our results and candidate materials. Sec.\,\ref{conclusion} summarizes this work.


%
%
\hypertarget{model-and-method}{%
\section{Model and Method}\label{model-and-method}}

\begin{figure}[h!]
\centering
\includegraphics[width=0.98\columnwidth]{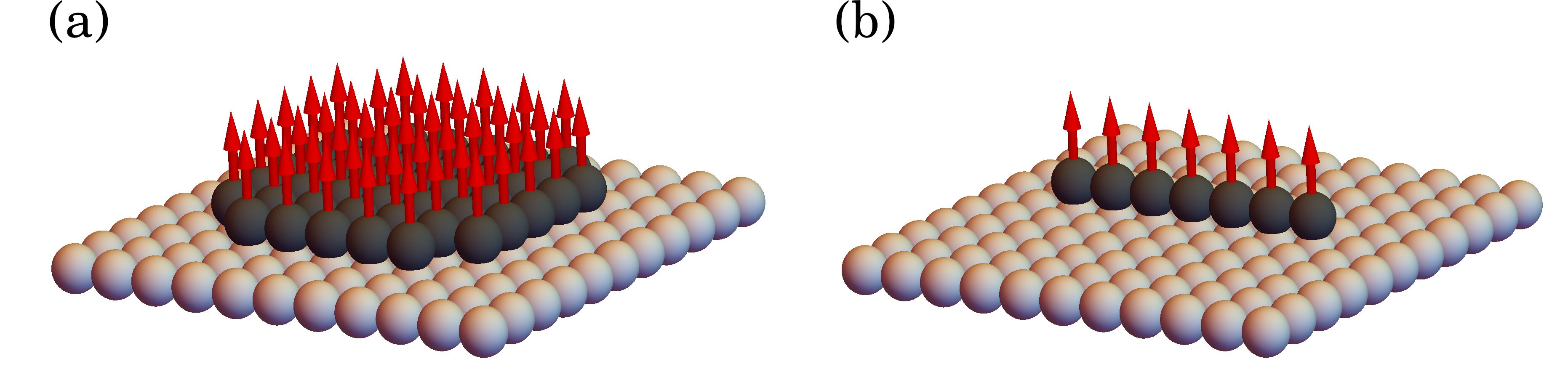}
\caption{Schematic setup of Shiba island (a) and Shiba chain (b) as studied in this paper.}
\label{fig:setup}
\end{figure}

We investigate the spectral and topological properties of magnetic
adatoms deposited on the surface of an unconventional superconductor.
The superconductor is modeled as a two-dimensional square lattice denoted by $\Lambda$ spanned by \(\hat{\vec{e}}_1\) and \(\hat{\vec{e}}_2\) and the magnetic adatoms occupy a subset $\Lambda^\star \subseteq \Lambda$. They are aligned ferromagnetically
and treated as classical spins. The tight-binding or Bogoliubov--de Gennes Hamiltonian is defined as
\begin{align}
\nonumber
H = &\sum_{\vec{r} \in \Lambda} \big[
t (c_\vec{r}^\dagger c_{\vec{r} + \hat{\vec{e}}_1} + c_\vec{r}^\dagger c_{\vec{r} + \hat{\vec{e}}_2} )
- \frac{\mu}{2} c_\vec{r}^\dagger c_\vec{r}
+\!\sum_{\vec r'  \in \Lambda}\Delta_{\vec r\vec r'}c_{\vec{r},\uparrow}^\dagger c_{\vec{r}',\downarrow}^\dagger \\
\nonumber
&+ i\alpha(c_\vec{r}^\dagger \sigma_2 c_{\vec{r} + \hat{\vec{e}}_1} - c_\vec{r}^\dagger \sigma_1 c_{\vec{r} + \hat{\vec{e}}_2})
+ {\rm H.c.} \big] \\
\label{eq:hamiltonian_rs}
&+ J \sum_{\vec{r} \in \Lambda^*} c_\vec{r}^\dagger \sigma_3 c_\vec{r}
\end{align}

Here
\(c_\vec{r}^\dagger = \begin{pmatrix} c_{\vec{r},\uparrow}^\dagger & c_{\vec{r},\downarrow}^\dagger \end{pmatrix}\)
is a spinor of the creation operators for electrons at site \(\vec{r}\)
with spin \(\uparrow,\downarrow\), and \(\sigma_{\alpha}\) are the three
Pauli matrices. \(t\) is the nearest-neighbour hopping amplitude;
\(\mu\) the chemical potential; \(\alpha\) the Rashba spin-orbit
coupling strength; and \(J\) the Zeeman strength resulting from the magnetic moments of the adatoms. Superconductivity is
induced by proximity effect, leading to the real-space pairing amplitude $\Delta_{\vec r\vec r'}$. 
For the case $\Lambda=\Lambda^\star$ with periodic boundary conditions (PBCs) imposed, the resulting
Bogoliubov--de Gennes Hamiltonian in momentum space reads
\begin{align}
H = &\frac{1}{2} \sum_\vec{k} \psi_\vec{k}^\dagger \mathcal{H}_\vec{k} \psi_\vec{k}\ , \\[5pt]
\nonumber
\mathcal{H}_\vec{k} = &[2t (\cos k_x + \cos k_y) - \mu] \tau_3 \otimes \sigma_0 \\[3pt]
\nonumber
& + 2 \alpha (\sin k_y \tau_3 \otimes \sigma_1 - \sin k_x \tau_0 \otimes \sigma_2) \\[3pt]
& + J\tau_3 \otimes \sigma_3 + \Delta_\vec{k} \tau_1 \otimes \sigma_1\ . 
\label{eq:hamiltonian}
\end{align}
Here
\(\psi_\vec{k} = \begin{pmatrix} c_{\vec{k}, \uparrow} & c_{\vec{k}, \downarrow} & c_{\vec{k}, \uparrow}^\dagger & c_{\vec{k}, \downarrow}^\dagger \end{pmatrix}^T\)
is the Nambu spinor of electron and hole creation and annihilation
operators, for momentum \(\vec{k}\) and spin \(\uparrow,\downarrow\).
\(\sigma_{0,1,2,3}\) (\(\tau_{0,1,2,3}\)) are the identity and Pauli matrices acting on spin (particle-hole)
space.
The energy dispersion of the Bogoliubov--de Gennes Hamiltonian
\eqref{eq:hamiltonian} is
\begin{widetext}
\begin{align}
\label{eq:general-spectra}
E(\vec{k}) = \pm \sqrt{
 J^2 + \Delta_\vec{k}^2 + \varepsilon_\vec{k}^2 + \alpha_\vec{k}^2
  \pm 2 \sqrt{J^2 (\Delta_\vec{k}^2 + \varepsilon_\vec{k}^2)
  + \varepsilon_\vec{k}^2 \alpha_\vec{k}^2
  }
}
\end{align}
\end{widetext}
with the definitions
\begin{align}
\varepsilon_{\vec{k}} &= 2t(\cos k_x + \cos k_y) - \mu\ , \\[3pt]
\alpha_{\vec{k}} &= 2\alpha \sqrt{\sin^2 k_x + \sin^2 k_y}\ .
\end{align}
The Fourier transformation of the pairing term is given by $\sum_{\vec r, \vec r'}\Delta_{\vec r\vec r'}c_{\vec r,\up}^\dag c_{\vec r',\dw} + {\rm H.c.} = \sum_{\vec k}\Delta_{\vec k} c_{\vec k,\up}^\dag c_{\vec k,\dw}^\dag + {\rm H.c.}$ with the superconducting order parameter (or gap function) $\Delta_{\vec k}$ in momentum space.
There are many ways of writing down the different \(s\)-wave pairings
\citep{wenger_d_1993} (we adopt the notation of Ref.\,\onlinecite{wenger_d_1993} in the following). 
The simplest case is the nodeless, isotropic onsite pairing,
\begin{equation}
s_{000}:\quad \Delta_{\vec k} = \Delta
\end{equation}
and the extended, \ie first-neighbor and second-neighbor $s$-wave pairings,
\begin{align}
	s_{100}:\quad \Delta_{\vec k} &= \Delta_{100} (\cos k_x + \cos k_y)\ , \\[3pt]
	s_{110}:\quad \Delta_{\vec k} &= 2\Delta_{110} \cos k_x \cos k_y\ .
\end{align}
as well as combinations thereof, \eg \(s_{000} + s_{100}\) etc. When we consider combinations, we always choose equal weight, \ie $\Delta = \Delta_{100} = \Delta_{110}$ and differently weighted combinations can be extrapolated.
We also study (chiral) combinations involving $d$-wave pairings $d+id \equiv d_{xy} + i d_{x^2-y^2}$ and $s+id\equiv s_{000} + i d_{xy}$ with
\begin{align}
d_{x^2-y^2}:\quad \Delta_{\vec k} &= \Delta (\cos k_x - \cos k_y)\ , \\[3pt]
d_{xy}:\quad \Delta_{\vec k} &= \Delta \sin k_x \sin k_y\ .
\end{align}

In Sec.\,\ref{results2D}, we follow the strategy to  first consider an infinitely large lattice entirely
covered by adatoms (\ie $\Lambda=\Lambda^*$, 
PBCs imposed) to simulate
the bulk properties of the systems. We derive a
topological phase diagram as a function of \(J\) and \(\mu\) by calculating the Chern number using the 
Fukui-Hatsugai-Suzuki (FHS) method\,\citep{fukui_chern_2005}. The Chern number is given by
\begin{equation}
\label{chern}
	\mathcal{C} = \frac{1}{2\pi} \int_{BZ} d\vec{k} \, F_{xy}(\vec{k})
\end{equation}
and \new{ $F_{xy}=\partial_{k_x} A_y(\vec k) - \partial_{k_y} A_x(\vec k)$ is the Berry curvature. $A_i = -i \sum \langle n, \vec k| \partial_{k_i}|n, \vec{k}\rangle$ is the corresponding Berry connection and $|n, \vec{k}\rangle$  the Bloch state (\ie eigenstate of the Bloch matrix) at momentum $\vec{k}$ of the $n$-th band.} In the FHS method\,\citep{fukui_chern_2005} one computes a discretized version of the Chern number,
\begin{equation}
	\mathcal{C}_n = \frac{1}{2 \pi i} \sum_l \tilde{F}_{12} (\vec{k}_l)\ ,
	\end{equation}
	\begin{equation}\nn
	\tilde{F}_{12} (\vec{k}) = \ln\big[U_1 (\vec{k})U_2(\vec{k} + \hat{\vec{1}})U_1(\vec{k} + \hat{\vec{2}})^{-1}U_2(\vec{k})^{-1}\big]
	\end{equation}
	where $U_1(\vec{k}) = \langle n, \vec{k} | n, \vec{k} + \hat{\vec{1}}\rangle$ \new{and $\hat{\vec{1}}$ ($\hat{\vec{2}}$) denotes the smallest discrete step in $k_x$ ($k_y$) direction.} While easy to implement and efficient to compute, for systems with band crossings or degeneracies the FHS method sometimes produces ambiguous results. Therefore we have double-checked large regions of our topological phase diagrams using the projector method\,\cite{prodan-10prl115501,palacio-morales_atomic-scale_2019} and verified that changes of the Chern number coincides with gap closing points.
On
identifying a suitable nontrivial phase, we then compute the spectra for
a cylinder geometry of the very same system and identify chiral edge modes at the ends of the
cylinder in agreement with the Chern number. We then switch to a more realistic real-space model with a
finite lattice with open boundary conditions (OBCs) and a circular island of adatoms embedded into the larger SC substrate. For this model
we compute the Lorentzian-smoothed LDOS directly (with Lorentzian broadening $\gamma$). In Sec.\,\ref{results1D} we repeat this
analysis for chains of adatoms. We directly compute the
relevant \(\mathbb{Z}_2\) topological invariant which is defined by 
\begin{equation}
\label{Z2}
	\nu = \text{sgn} (\text{Pf}[iH(0)]\text{Pf}[iH(\pi)])
\end{equation}
where $\text{Pf}[ \cdot ]$ is the Pfaffian and $H(k)$ the Hamiltonian matrix at momentum $k$.
We complement these results with OBC spectra as a function of $\mu$. Eventually we consider the experimentally relevant system of a chain embedded into a 2D substrate. The latter system can be different from the purely 1D case.


The derived phase diagrams are always complemented by calculations of the gap size and, in particular, gap closing lines or areas (only shown when significant). This allows us to independently verify the topological phase diagrams, since changes of the topological invariant are only possible when the bulk gap closes.
Gap closing points are found from the roots of the
dispersion in \autoref{eq:general-spectra}. These are found when either
\(\alpha_\vec{k}\) or \(\Delta_\vec{k}\) vanish, which occur at points
of high symmetry. For extended \(s\)-wave this corresponds not just to
lines of gap closing but also to whole hyperbola bounded regions in
parameter space. These regions are bounded by the functions
\(\varepsilon_\vec{k}^2 = J^2 - \Delta_\vec{k}^2\) evaluated at
\(\vec{k} = (0,0), (\pi,\pi), (0, \pi), (\pi,0)\).


For extended $s$-wave superconductors with Rashba spin-orbit coupling, Zhang \ea~predicted in 2013 a time-reversal invariant (TRI) TSC phase\,\citep{zhang_time-reversal-invariant_2013}. This phase is realized when the Fermi surfaces encircle an odd number of TRI momenta and the superconducting gap function changes its sign between these Fermi surfaces. In our phase diagrams, we identify
several of these phases (in 1D and 2D) at $J=0$ by examining the Fermi surfaces,
and by confirming the presence of edge modes in cylinder spectra or real
space LDOS. These helical Majorana modes persist even for $J\not= 0$ as long as the bulk gap remains finite although the modes are no longer protected. In case of the Shiba chains, these TRI TSC phases do not appear when the chain is deposited onto a larger two-dimensional substrate. We demonstrate, however, that we find this phase for anti-Shiba chains, \ie a 2D area covered by magnetic atoms where a chain of atoms is missing. For this system it is possible to identify parameters where the TRI TSC phase is stabilized, although the surrounding substrate is magnetic and thus breaks time-reversal symmetry. We show that the helical Majorana bound states even survive in the presence of random disorder.

%
%

\hypertarget{results}{%
\section{Results for Shiba islands}\label{results2D}}


In this section, we present results for various unconventional substrates. We always show the topological phase diagram where colorful areas correspond to a finite Chern number as indicated by the color bar. White regions are topologically trivial with Chern number $C=0$, while grey regions mark gapless (and thus non-topological) areas. Red regions, usually centered around $J=0$ lines, label TRI topological phases of the Zhang-Kane-Mele type (by definition they must have $C=0$) which are discussed separately in Sec.\,\ref{tri-phases}. Dashed lines and crossings of dashed lines which are shown within the phase diagrams correspond to the selected parameters for the  spectral plots. 

These spectral plots are derived for cylinder and real space geometries. Cylinders usually contain 100 unit cells in the finite direction. We preferably choose parameters with $|C|=1$ and thus only a single chiral edge mode (per edge). Edge modes on different edges are shown in blue and red, respectively, while bulk states appear in black. Cylinder spectra of the TRI phases possess right and left movers on both edges; they are thus shown in maroon. Where required, we also show the spatial profile of zero-energy edge states of the cylinder spectra.

Real space plots reveal the spatial LDOS at zero energy (the region covered by magnetic adatoms is shown as black dots). In addition, we also show the energy-dependent LDOS measured at a lattice site located at the boundary of the Shiba island or chain. Energy peaks within the topological gaps are shown in red while peaks associated with bulk states in black.

We start with a conventional, nodeless substrate to connect our results to the existing literature, before we discuss $d$-wave, chiral singlet ($d+id$ and $s+id$) and various extended $s$-wave substrates. Finally we focus on the TRI topological phases.

%
%
\subsection{Nodeless $s$-wave substrate}\label{s_000-s-wave}

\begin{figure}[t!]
\centering
\includegraphics{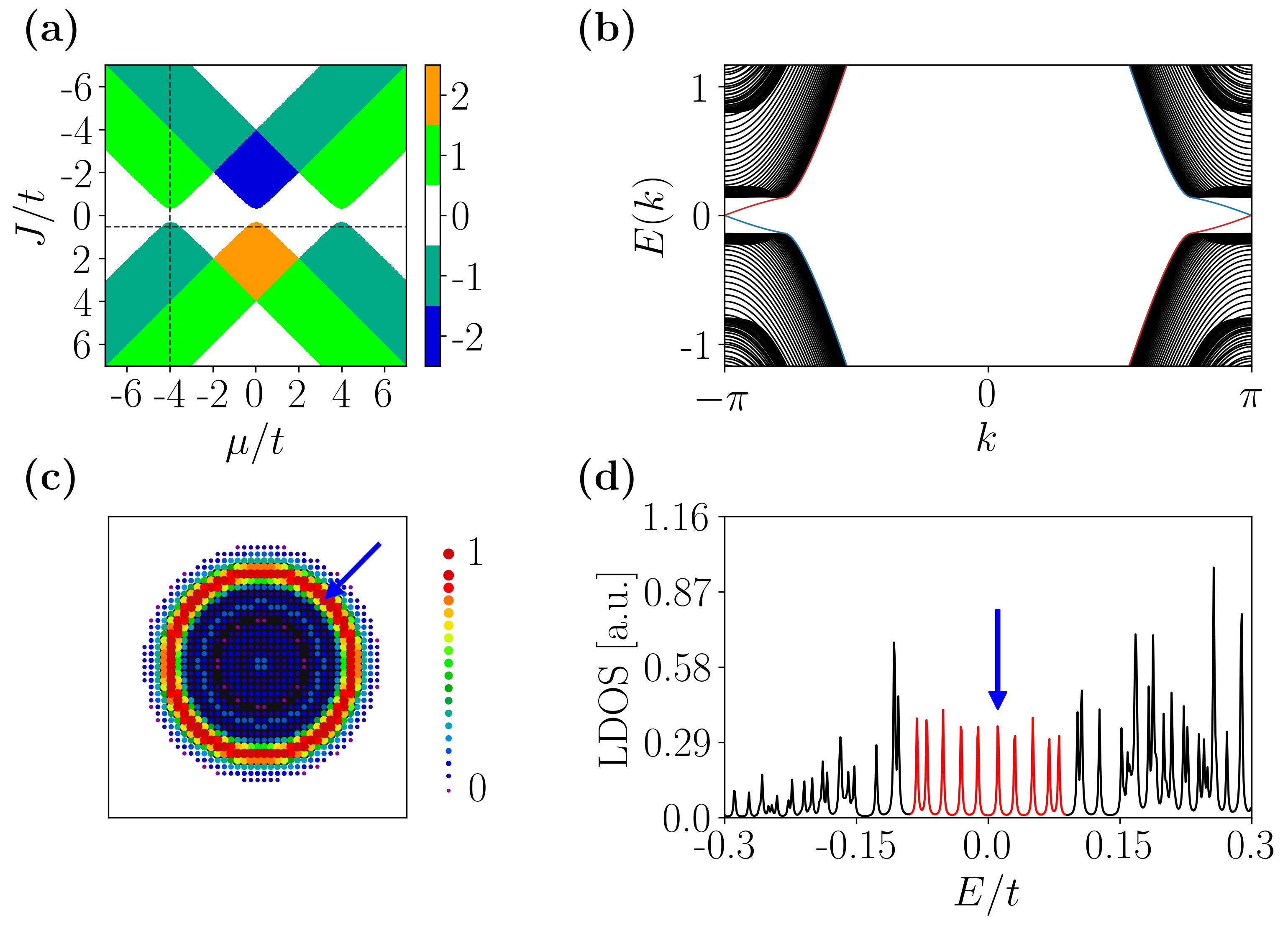}
\caption{Isotropic, nodeless $s$-wave substrate. (a) Topological phase diagram. (b) Cylinder spectrum in the $C=-1$ phase. (c) Real space LDOS for the energy state closest to $E=0$, as indicated by the arrow in panel (d). (d) Energy-resolved LDOS  measured at the edge of the island.
Parameters used in (b)-(d):
$\Delta = 0.3t, \mu = -4t, J = 0.5t, \alpha=0.2t$, and Lorentzian broadening $\gamma = 0.001t$.}
\label{fig:s-wave}
\end{figure}

Here we briefly show the simple, nodeless $s$-wave substrate, $s_{000}$ in the notation of Ref.\,\onlinecite{wenger_d_1993}, as studied before\,\cite{rachel_quantized_2017} and as realized in both Fe/Pb(110)\,\cite{nadj-perge_observation_2014}, Co/Pb(110)\,\cite{menard_two-dimensional_2017} and Fe/Re(0001) systems\,\cite{kim_toward_2018}. For reference only we
present the results in \autoref{fig:s-wave}. Topologically non-trivial phases with Chern numbers
\(C=0, \pm 1, \pm 2\) are easily accessible, and the whole parameter space
is well gapped. Hence edge modes are readily found.

%
%
\subsection{$d$-wave substrate}\label{d-wave}

\begin{figure}[b!]
\centering
\includegraphics{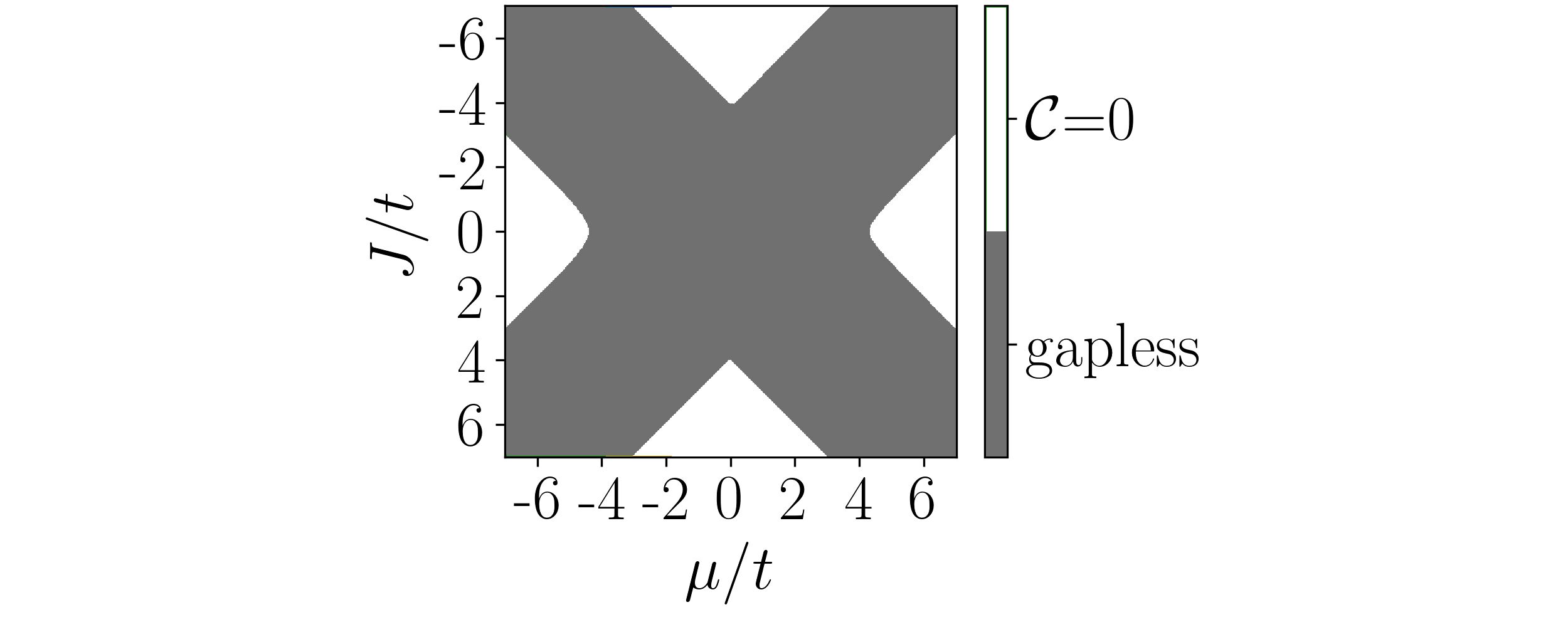} 
\caption{
Nodal \(d_{x^2-y^2}\) and \(d_{xy}\)-wave substrate. The phase diagram
only contains gapped $C=0$ phases (white) and gapless regions (grey).}
\label{fig:dwave-gaps-cherns}
\end{figure}

There are two different $d$-wave pairings, $d_{x^2-y^2}$ and $d_{xy}$, and the former is the one which is realized in most of the cuprate high-temperature superconductors.
Both exhibit
identical gap closing points and topological phases, see Fig.\,\ref{fig:dwave-gaps-cherns}. This can be readily understood from the nodal lines of their superconducting order parameters which are related by a
\(\pi/4\) rotation. It turns out that their gapless points in the spectra persist over a very large range of parameters.
Gapped phases are accessible only at unreasonably large
chemical potential (\eg when the chemical potential is below or above the normal band structure) or Zeeman fields leading to trivial phases. The gaplessness is easily understood: in the absence of Rashba spin orbit coupling and Zeeman field, the superconducting spectrum is given by $\sqrt{\eps_{\vec k}^2 + \Delta_{\vec k}^2}$. Simultaneous zeros in $\eps_{\vec k}$ and $\Delta_{\vec k}$ lead to gapless points, \ie the crossings of the nodal lines of the superconducting order parameter with the FS are responsible for the gaplessness. Additional Rashba and Zeeman terms do not lead to qualitative changes.

\begin{figure}[h!]
\centering
\includegraphics{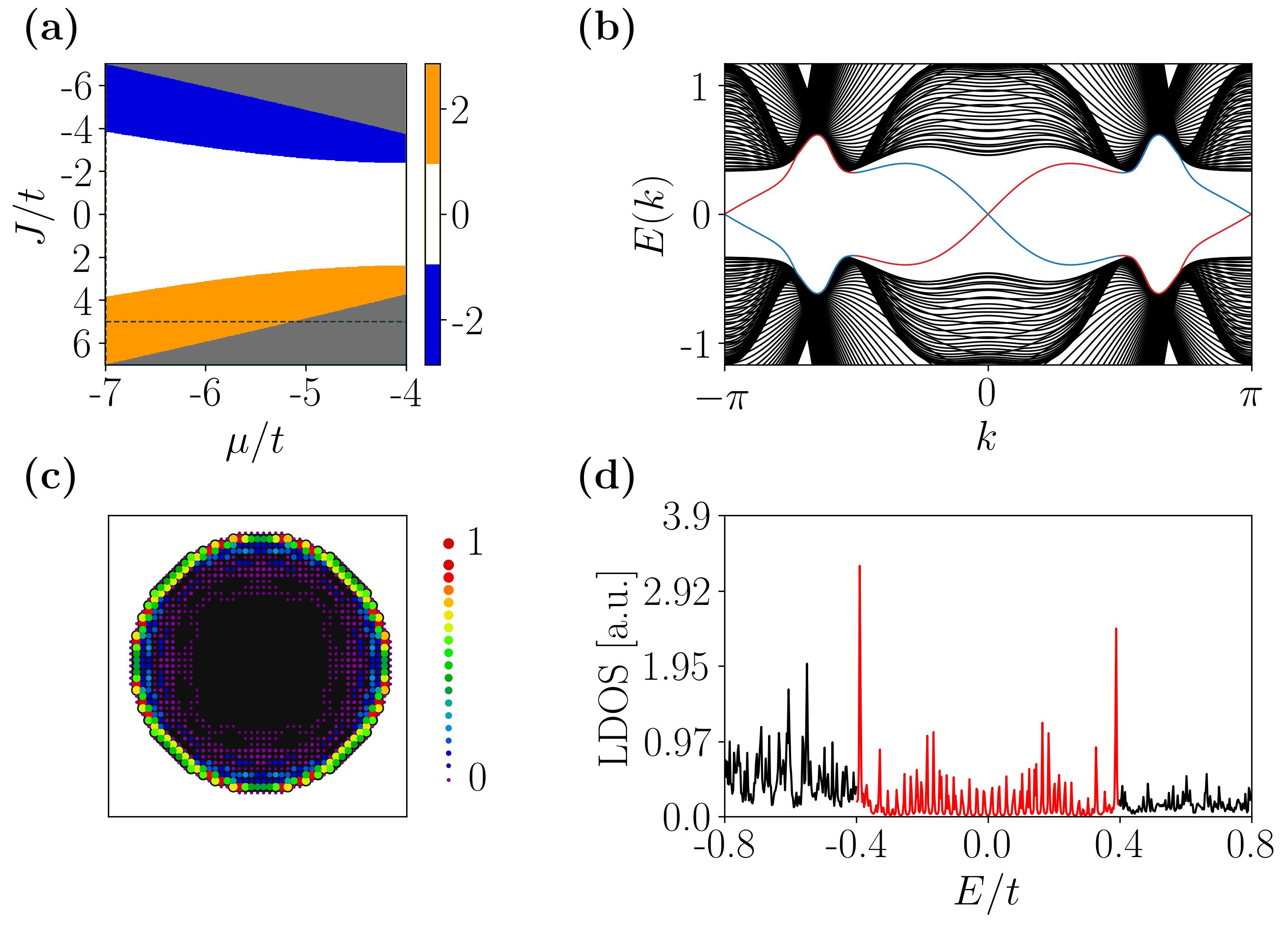}
\caption{Nodal $d_{x^2-y^2}$-wave substrate for a normal state system with strong second-neighbor hoppings $t_2=t$. (a) Topological phase diagram \new{with gapless regions in grey.} (b) Cylinder spectrum. (c) Real space LDOS for the energy closest to $E=0$. (d) Energy-resolved LDOS measured at an edge site of the island. Parameters used in (b)-(d):
$t_2=t, \alpha = 0.8t, \Delta = 1.2t, \mu = -7t, J = 5t, \gamma = 0.001t$.}
\label{fig:d-wave-nnn}
\end{figure}

In the following, we assume that we could significantly manipulate the FS of the normal state system by applying stress, strain or pressure with the aim to avoid crossing of the FS with the SC order parameter. This can be achieved, for instance, by significant next-nearest-neighbor hoppings $t_2 > 0.7 t$. Such parameters lead to additional gapped phases with finite Chern numbers. We show in Fig.\,\ref{fig:d-wave-nnn} an example for $t_2=t$. These results suggest that, in principle, we can find topologically non-trivial phases in MSH structures with $d$-wave substrates --- provided we find ways to avoid gap closings.

%
%
\subsection{Chiral singlet substrates}\label{did}

\begin{figure}[b!]
\centering
\includegraphics{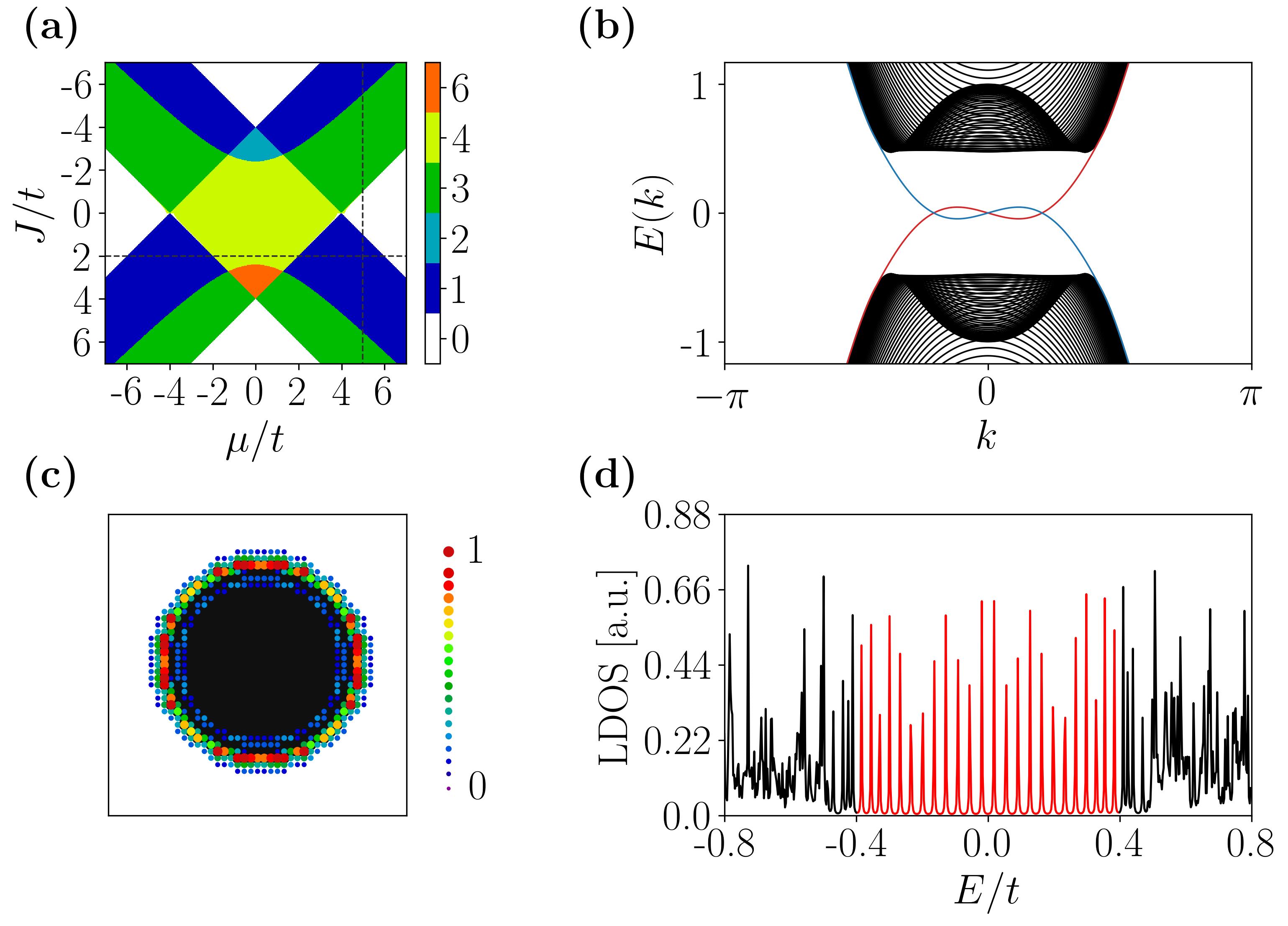}
\caption{
Nodeless, chiral $d+id$-wave substrate. (a) Topological phase diagram. (b) Cylinder spectrum. (c) Real space LDOS for the energy closest to $E=0$. (d) Energy-resolved LDOS measured at an edge site of the island. Parameters used in (b)-(d): $\alpha = 0.8t, \Delta = 1.2t, \mu = 5t, J = 2t, \gamma = 0.001t$.}
\label{fig:d-id}
\end{figure}

The resulting topological superconducting phases of the considered MSH structures are characterized by a finite Chern number, thus they constitute an example of chiral, \ie time-reversal broken, superconductivity. Chiral superconductivity can also emerge as an intrinsic, unconventional pairing. Sometimes, competing superconducting instabilities can become energetically comparable, either due to fine-tuning of 
\begin{figure}[t!]
\centering
\includegraphics{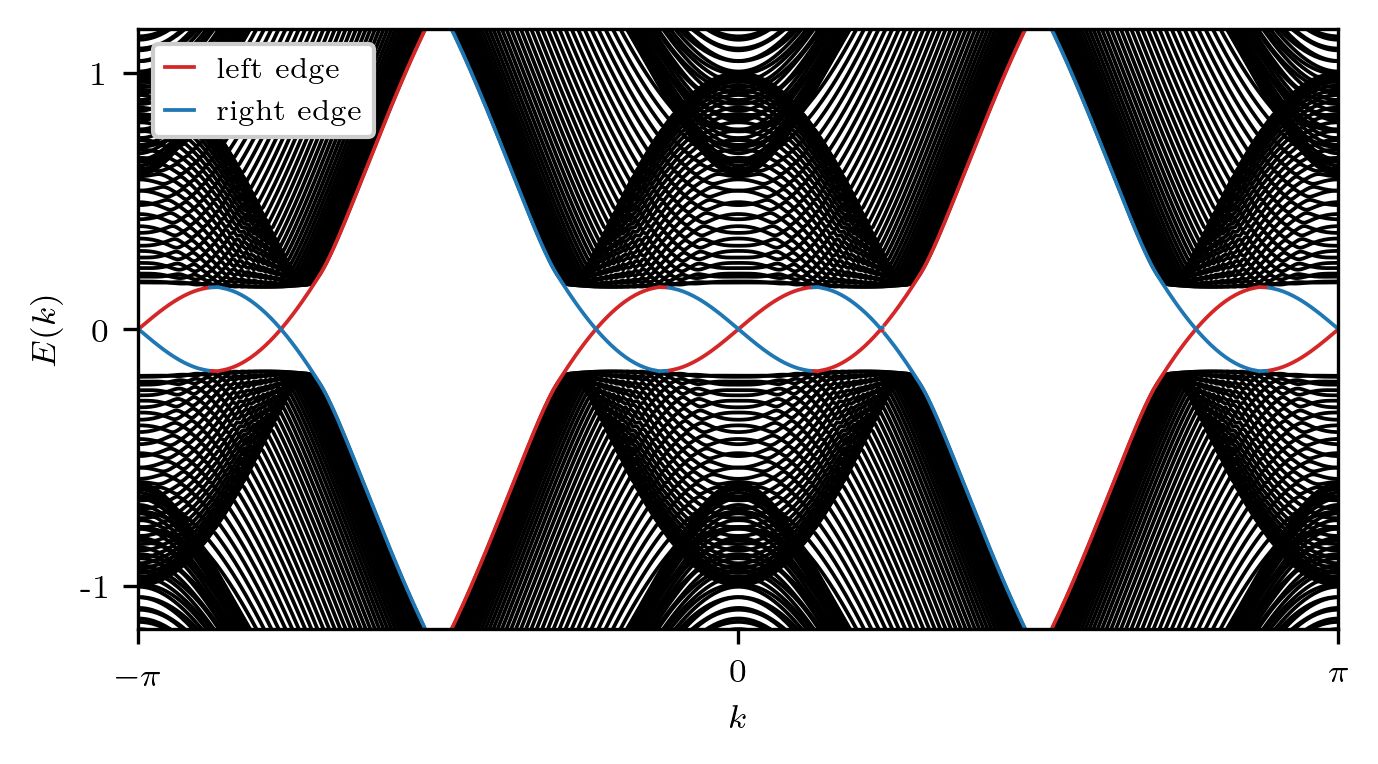}
\caption{Cylinder spectrum for the $C = 6$ phase for a MSH structure with $d+id$-wave substrate.
Note the corresponding six chiral states on either edge. Parameters used:
$\alpha = 0.8t, \Delta = 1.2t, \mu = 0t, J = 3t$.}
\label{fig:d-id_cylinder}
\end{figure}
\begin{figure}[b!]
\centering
\includegraphics{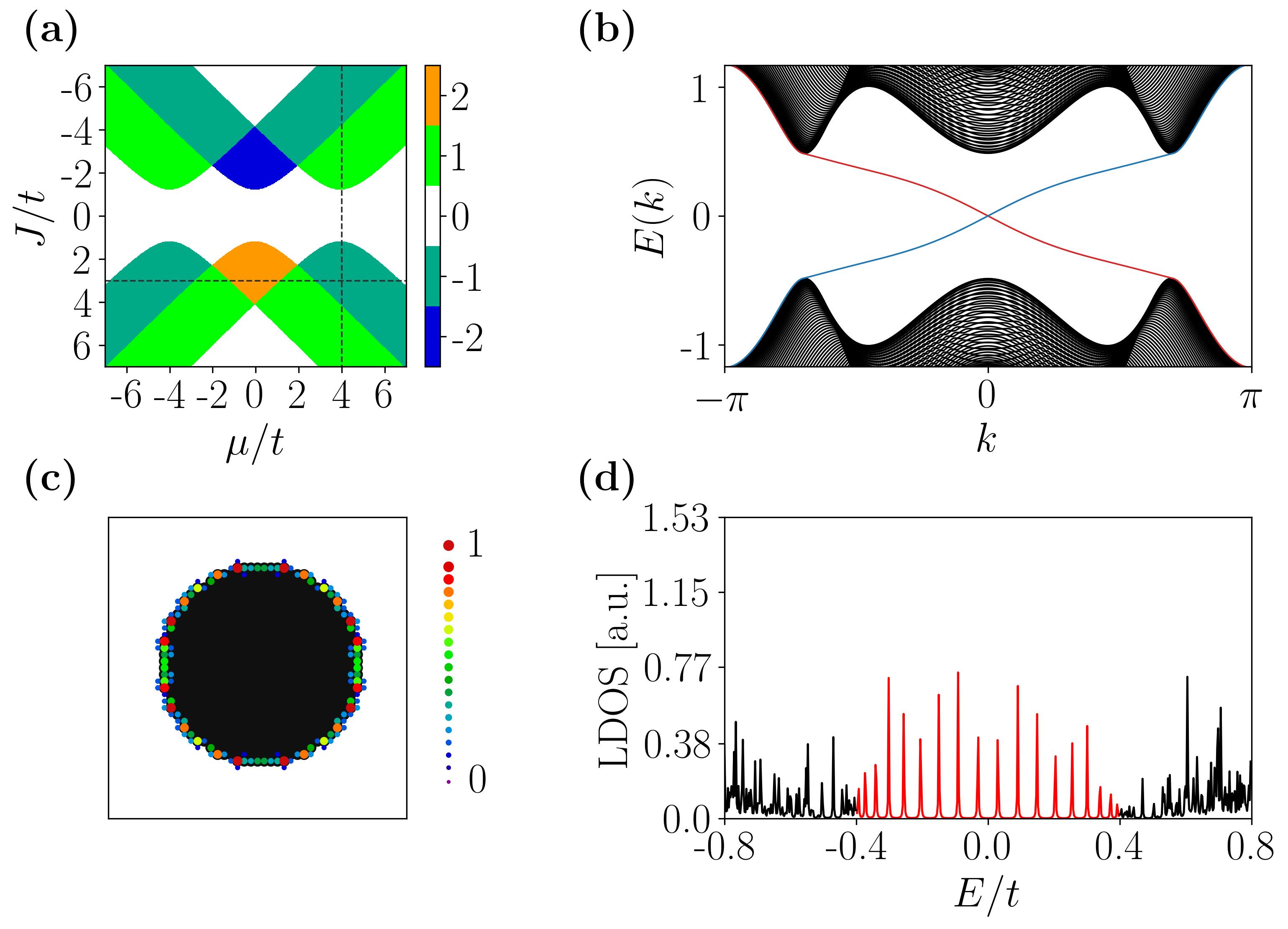}
\caption{
Nodeless, chiral $s+id$-wave substrate. (a) Topological phase diagram. (b) Cylinder spectrum. (c) Real space LDOS for the energy closest to $E=0$. (d) Energy-resolved LDOS measured at an edge site of the island. Parameters used in (b)-(d):
$\alpha = 0.8t, \Delta = 1.2t, \mu = 4t, J = 3t, \gamma = 0.001$.}
\label{fig:s-id}
\end{figure}
\begin{figure*}[t!]
\centering
\includegraphics{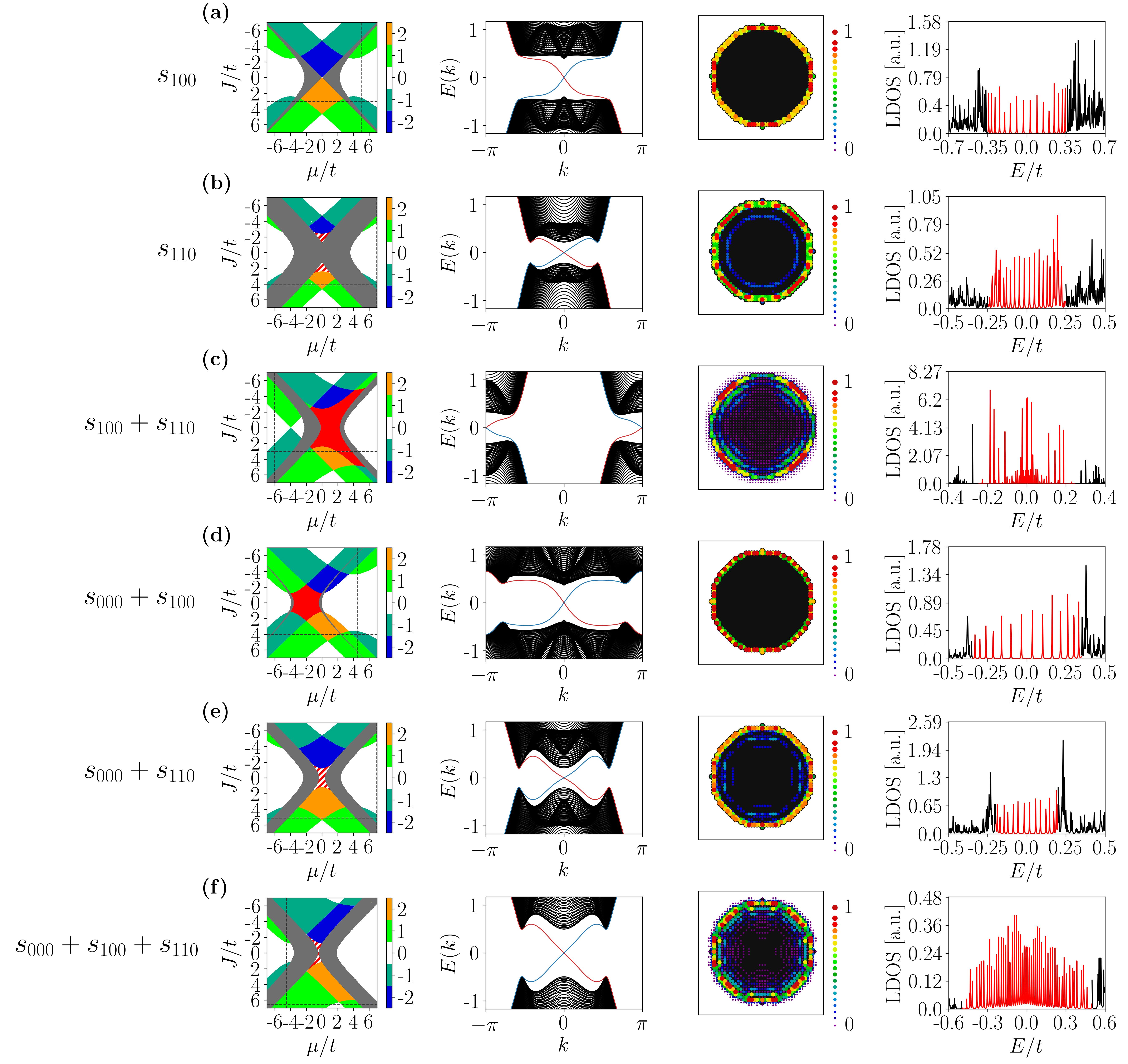}
\caption{
Extended \(s\)-wave substrates. (From left to right): Topological phase diagram, cylinder spectrum, real space LDOS for the energy closest to $E=0$, and energy-resolved LDOS measured at an edge site of the island. 
(a) Nearest-neighbor pairing $s_{100}$. Parameters used: $\mu=5t, J=3t, \gamma = 0.001$.
(b) Next-nearest neighbor pairing $s_{110} \equiv s^\pm$. Parameters used: $\mu=6.9t, J=4.1t, \gamma=0.001$.
(c) Combination of $s_{100}$ and $s_{110}$. Parameters used: $\mu=-6t, J=3t, \gamma=0.0001$.
(d) Combination of $s_{000}$ and $s_{100}$. Parameters used: $\mu=4.5t, J=4t, \gamma=0.001$.
(e) Combination of $s_{000}$ and $s_{110}$. Parameters used: $\mu=6.9t, J=5.1t, \gamma=0.001$.
(f) Combination of $s_{000}$, $s_{100}$ and $s_{110}$. Parameters used: $\mu=-4.5t, J=6.5t, \gamma= 0.001$.
Parameters used for all panels: $\alpha=0.8t, \Delta = 1.2t$.
In (c)-(f) we have chosen equal amplitudes of the combined pairing potentials, \eg $\Delta_{100} = \Delta_{110}\equiv\Delta$. \new{TRI TSC phase is shown in red, $C=0$ phases with trivial edge states shown in white-red striped.}
}
\label{fig:ext_s-wave}
\end{figure*}
some external parameter or when a superconducting instability is associated with a two-dimensional irreducible representation. In the latter case, the degeneracy is stable as long as the lattice symmetry remains intact. Quantum mechanical degeneracies allow for arbitrary superpositions, and it turns out that the complex superposition $x\pm i y$ for representations $x$ and $y$ usually maximizes the condensation energy. While energetically convenient, nature has to choose between a chirality, \ie between $x+iy$ and $x-iy$, and thus spontaneously breaks time-reversal symmetry. Moreover, the complex superposition of order parameters leads to a {\it full} gap. Chiral superconductors can be topologically non-trivial (\eg for $d+id$) or trivial (\eg $s+id$), and they can be fine-tuned from one phase to another. Although the chiral $d+id$-wave superconductor is already topologically non-trivial ($C=2$) and no further ingredients are required, we nevertheless discuss it here as an MSH structure. Note that the PBC phase
diagrams Fig.\,\ref{fig:d-id}\,(a) and Fig.\,\ref{fig:s-id}\,(a) were previously reported in Ref.\,\onlinecite{varona_topological_2018}.

Results for the MSH system with $d_{x^2-y^2} + i d_{xy}$-wave substrate are presented in Fig.\,\ref{fig:d-id}.
We observe
a topologically rich system. In particular, the $C=2$ phase at $\alpha=J=\mu=0$ immediately turns into a $C=4$ phase for any finite Rashba spin-orbit coupling present. Amongst the topological phases, we note that even a $\mathcal{C} = 6$ phase is present. For the interested reader, we show the cylinder spectrum for a parameter point within this phase in Fig.\,\ref{fig:d-id_cylinder}; indeed one can nicely count six chiral edge modes (per edge).

The $s_{000}+i d_{xy}$-wave ($s+id$-wave in the following) superconductor is also an inherent, chiral superconductor.
It is topologically trivial with $C=0$, yet it breaks TR symmetry and is nodeless.
As mentioned above, changing parameters (or adding terms such as Rashba SOC or a Zeeman field) can drive the system into a topologically trivial phase. This is by no means surprising; in fact, the situation is analogous to topological band insulators where it is well-established that the ground state can be tuned from topologically non-trivial to trivial (and vice versa) upon changing a parameter, while the symmetry of the ground state does not change.

Results for the $s+id$ substrate are summarized in Fig.\,\ref{fig:s-id}. It is less topologically
rich than the $d+id$ case and is almost identical to the plain $s$-wave system discussed in Sec.\,\ref{s_000-s-wave}. Note that other chiral combinations of $s$ and $d$-wave pairings lead to very similar phase diagrams; for instance, the MSH structure with $s_{100}+id_{xy}$-wave substrate is more or less the same but the $C=2$ and $C=-2$ phases touch each other at $J=\mu=0$.

\begin{figure*}
\centering
\includegraphics{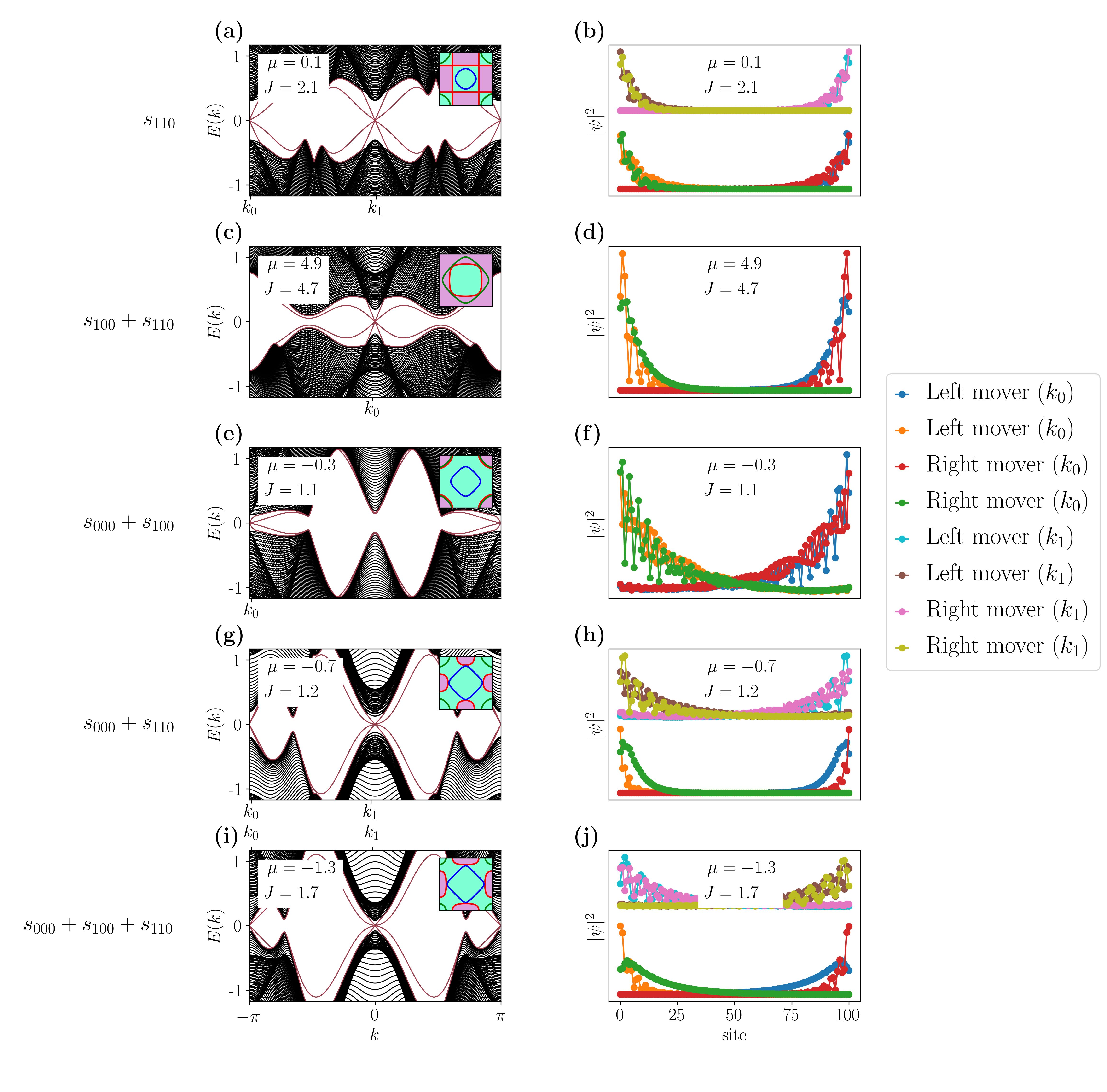}
\caption{
Cylinder spectra and wavefunctions for $C=0$ phases hosting edge states.
The five rows correspond to the $C=0$ phases with edge modes, shown in Fig.\,\ref{fig:ext_s-wave} as red or white-red striped regions.
Left column: cylinder spectra (cylinder length is 100 sites) which clearly reveal the edge states in maroon, bulk bands in black. Insets show the specific parameters as well as the Fermi surface(s) and the superconducting order parameter.
Right column: Wavefunctions $|\psi|^2$ of the four energy states closest to $E(k)=0$ at momentum $k=k_0$ and, in case of the trivial phases, also for $k=k_1$.
Panels (c) and (e) correspond to the TRI topological superconductor with {\it one} pair of helical edge modes per edge (see panels (d) and (f)), while panels (a), (g) and (i) are topologically trivial (with {\it two} pairs of helical edge modes per edge (see panels (b), (h) and (j)).
Parameters used for all panels: \(\alpha=0.8t, \Delta = 1.2t\).}
\label{fig:ext_s-wave_tri}
\end{figure*}

%
%
\subsection{Extended $s$-wave substrates}\label{extended-s-wave}

We distinguish between the isotropic, nodeless $s$-wave pairing, formally expressed as an onsite pairing term in the Hamiltonian, and the ``dispersive'' nodal $s$-wave pairing, caused by pairing between nearest or further neighbor sites. The latter is broadly called {\it extended} $s$-wave pairing. Note that the symmetry of these superconducting order parameters remains the same; local and extended $s$-wave both realize the trivial representation. Extended $s$-wave pairing has attracted significant attention in the past decade because most of the iron-based superconductors realize the so-called $s^\pm$-pairing (in our notation, $s_{110}$). Here we study (next) nearest-neighbor pairing referred to as $s_{100}$ ($s_{110}$) and all combinations \new{of $s_{000}$, $s_{100}$, and $s_{110}$.}

All extended \(s\)-wave results are compiled and presented in
Fig.\,\ref{fig:ext_s-wave}. There are clearly similarities between some of the phase
diagrams and that of plain \(s_{000}\)-wave. As mentioned before, gapless regions are shown in grey. 
All phase diagrams contain gapless regions. While the combination of onsite and nearest-neighbor pairing (\ie $s_{000}+s_{100}$) leads to the smallest gapless region, pure nearest-neighbor pairing (\ie $s_{110}$) to the largest one. As for the isotropic $s$-wave case $s_{000}$, topological phases carry Chern numbers $C=0, \pm 1, \pm 2$. What changes, however, is the position and size of these topological phases in the phase diagram.
The most drastic difference to other phase diagrams is the presence of phases with $C=0$ which nevertheless features edge states; these phases are shown either in red or as red-white striped regions. Red phases are topologically non-trivial and realize time-reversal invariant (TRI) topological superconductivity as discovered by Zhang \ea\ in 2013; we will discuss this type of superconductivity in the next section. The red-white striped phases are topologically trivial phases with edge states. They feature an {\it even} number of pairs of helical edge states, thus leading to a topologically trivial phase. 

As mentioned before, \(s_{110}\) pairing is believed to be realized in many of the iron-pnictide and iron-chalcogenide superconductors including the material FeSe$_{0.5}$Te$_{0.5}$\,\cite{wang_evidence_2018, zhang_observation_2018} which has attracted much attention lately. Our results can thus be seen as a one-band toy model for MSH structures on FeSe$_{0.5}$Te$_{0.5}$ surfaces.

The results shown in Fig.\,\ref{fig:ext_s-wave} contain again cylinder spectra and spatial and energy-resolved LDOS plots. For all shown cases, we have selected the $C=-1$ phase featuring one chiral edge mode. There are clearly model dependent details, \eg the penetration length of the edge modes. These differences are, however, attributed to differences in gap size etc. Otherwise, the qualitative behavior in all models in roughly the same, as expected for topological phases.


%
%
\hypertarget{tri-phases}{%
\subsection{TRI phases}\label{tri-phases}}

All extended \(s\)-wave pairings except for \(s_{100}\) exhibit a TRI
TSC or trivial edge state phase in 2D (see phase diagrams in Fig.\,\ref{fig:ext_s-wave}).
\(s_{000}+s_{100}\) and \(s_{100}+s_{110}\) have a TRI TSC phase which is characterized by the $\mathbb{Z}_2$ invariant; the Chern number must be zero due to TR symmetry. This
can be seen by inspection of the involved Fermi surfaces (shown as insets in Fig.\,\ref{fig:ext_s-wave_tri}). \(s_{110}\),
\(s_{000}+s_{110}\), \(s_{000}+s_{100}+s_{110}\) have a trivial edge
state phase. The latter is identified because the Fermi surface is not
split by the order parameter and there are two zero-energy crossings
which by Kramer's theorem results in a trivial phase. 
Or with other words, there are two pairs of helical edge modes per edge. This is demonstrated in Fig.\,\ref{fig:ext_s-wave_tri} (b), (h), (j).
In either case,
these phases are associated with zero-energy states --- whether or not they are topologically protected experimentalists might detect them. This raises the question how to distinguish them in an experiment. Moreover, it remains unclear whether such trivial edge states could be of technological use of any kind. These two questions will be left for future work.

%
%
\hypertarget{shiba-chains}{%
\section{Results for Shiba chains}\label{results1D}}

Shiba chains have attracted much attention as a direct realization of the Kitaev chain. Such one-dimensional topological superconductors host zero-energy Majorana bound states at the chain end. Given the enormous interest, mainly because they can in principle be combined into wire networks as a device for topological braiding, here we extend our investigation to 1D MSH structures with unconventional substrates.

Shiba chains are conveniently described by the Hamiltonian \eqref{eq:hamiltonian_rs} by restricting the adatom lattice $\Lambda^\star$ to 1D. We will consider both that the substrate lattice $\Lambda$ is one-dimensional but also that it is two-dimensional (as illustrated in Fig.\,\ref{fig:setup}\,(b)). In particular, in Sec.\,\ref{sec:chain-in-2D}, we will consider the experimentally relevant case of a chain of magnetic adatoms embedded into a 2D superconducting substrate more thoroughly.
\begin{figure}[t!]
\centering
\includegraphics{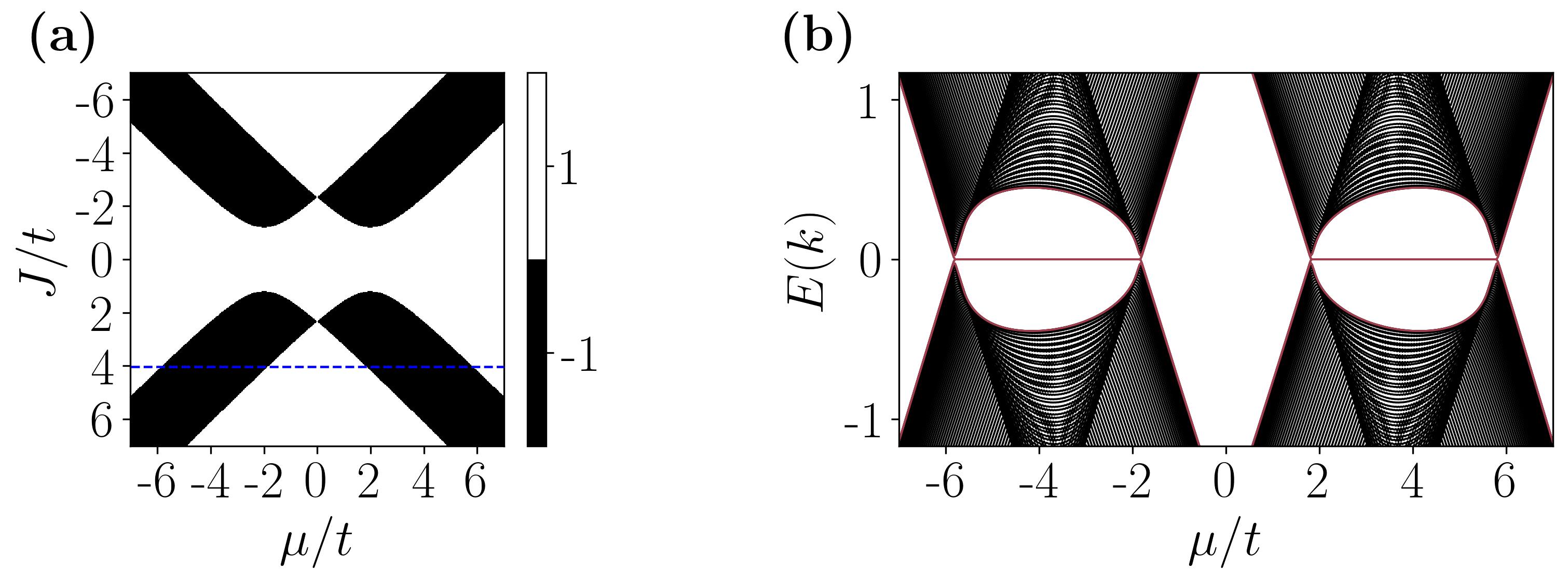}
\caption{Isotropic, nodeless s-wave substrate. (a) Topological phase diagram (white = trivial, black = topological). (b) Energy spectrum as a function of $\mu$ for $J=4t$ (indicated by blue line in (a)). Parameters used: $\alpha=0.8$, $\Delta=1.2$.}
\label{fig:s000-chain}
\end{figure}
\begin{figure}[t!]
\centering
\includegraphics{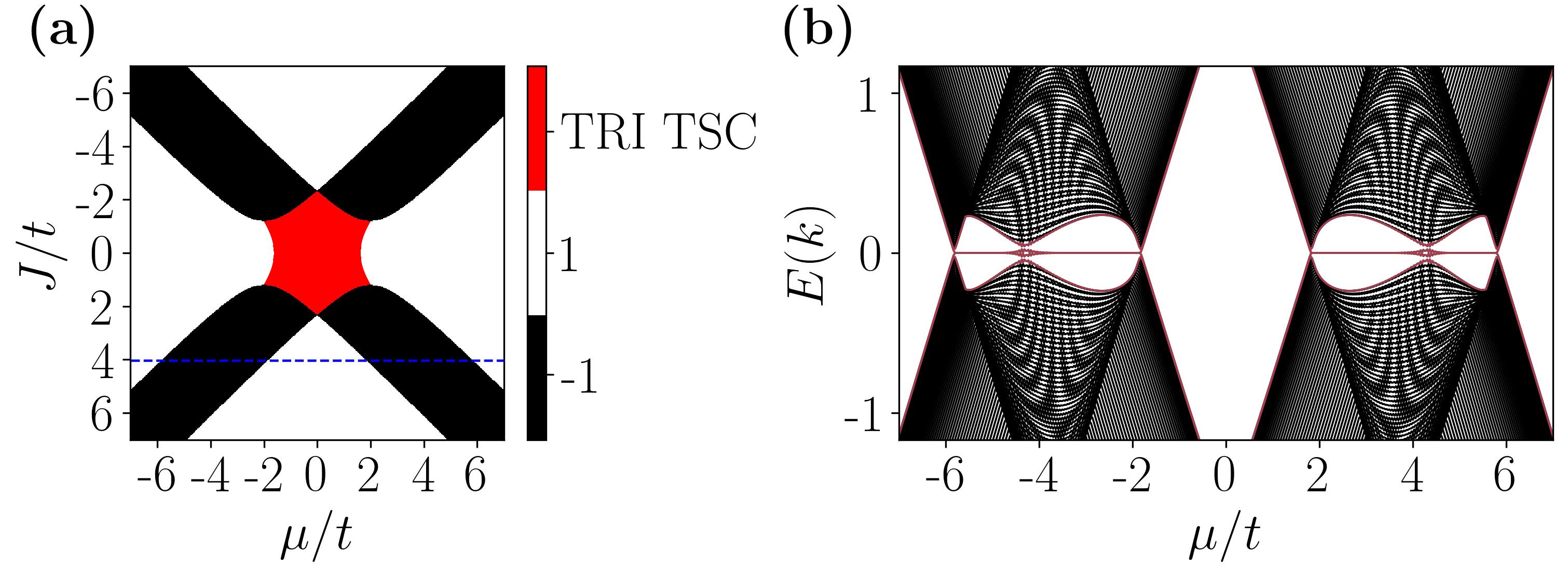}
\caption{
Extended $s_{100}$-wave substrate. (a) Topological phase diagram (white = trivial, black = topological, red = TRI topological). (b) Energy spectrum as a function of $\mu$ for $J=4t$ (indicated by blue line in (a)). 
Parameters used: $\alpha=0.8$, $\Delta=1.2$.
}
\label{fig:s100-chain}
\end{figure}
\begin{figure}[t!]
\centering
\includegraphics{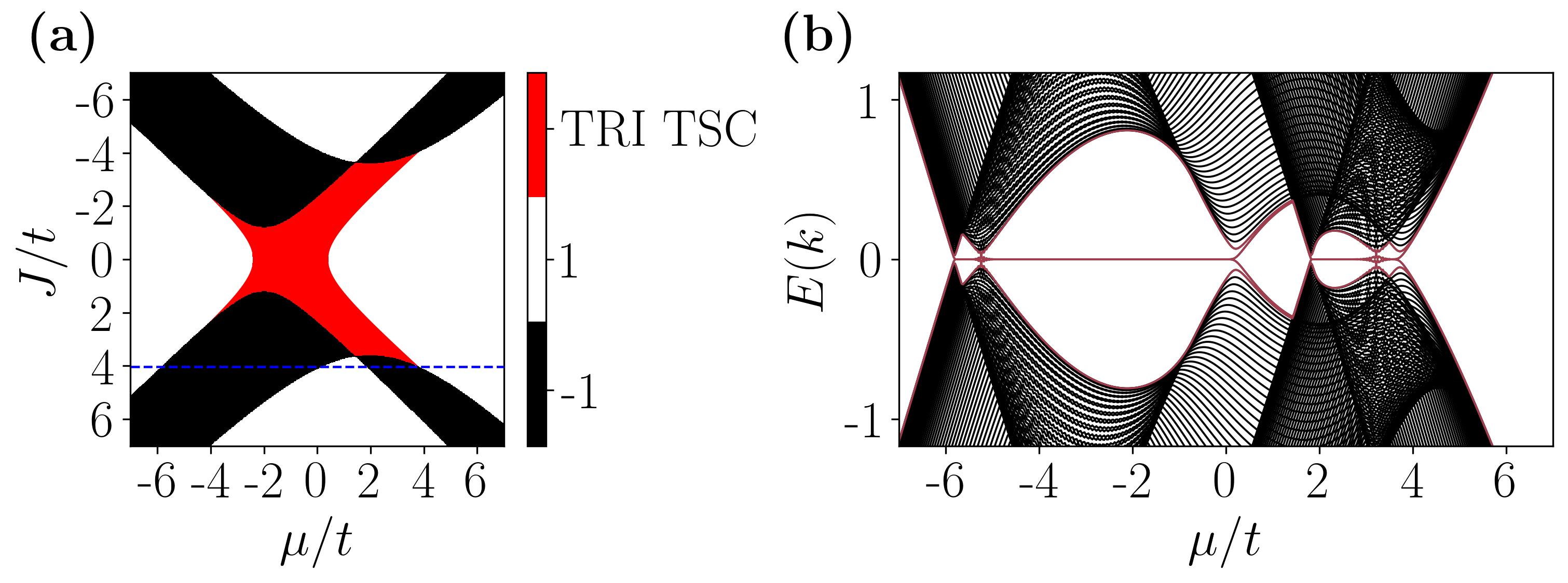}
\caption{
Extended $s$-wave substrate with $s_{000}+s_{110}$. (a) Topological phase diagram (white = trivial, black = topological, red = TRI topological). (b) Energy spectrum as a function of $\mu$ for $J=4t$ (indicated by blue line in (a)). Parameters used: $\alpha=0.8$, $\Delta=1.2$.}
\label{fig:s000_s110-chain}
\end{figure}

By restricting \new{$\Lambda$ and $\Lambda^*$ to be one-dimensional and }arranged along the $x$ direction, the superconducting gap functions reduce to  finite $k_x$, $\Delta(\vec k)=\Delta(k_x, 0)$. On doing this note that
the \(s_{100}, s_{110}, s_{100} + s_{110}\) and \(d_{x^2 - y^2}\)
pairings are  the same up to a constant factor.
Hence we show first the $s_{000}$ case for reference (see Fig.\,\ref{fig:s000-chain}), and then discuss the extended $s$-wave substrates with $s_{100}$ and $s_{000}+s_{100}$ (see Figs.\,\ref{fig:s100-chain} and \ref{fig:s000_s110-chain}). We show for all the three cases the topological phase diagram as a function of $J$ and $\mu$ obtained for PBCs. For the purely 1D case we directly computed the topological $\mathbb{Z}_2$ invariant which is 1 ($-1$) in the trivial (topological) phase. Then we show the energy spectrum for OBC as a function of $\mu$; zero-energy states in otherwise gapped regions are shown in maroon, to emphasize the topological phase. 

\new{Extended $s$-wave pairings do not lead to drastic changes of the phase diagram compared to the isotropic $s_{000}$ substrate. Most notably, we find additionally the TRI TSC phase of Zhang--Kane--Mele\,\cite{zhang_time-reversal-invariant_2013}. As in 2D, the phase extends the $J=0$ line to finite $J$ (see discussion below). By combining isotropic $s_{000}$ and extended $s_{110}$ pairing [see Fig.\,\ref{fig:s000_s110-chain}] the $\mu \to -\mu$ symmetry is broken. Regardless of what $s$-wave substrate we consider, there are always extended regions in the phase diagram where the topological superconducting phase is realized and MZMs are present.}

%
%
\subsection{Shiba chains on a 2D substrate revisited}\label{sec:chain-in-2D}

After studying 2D Shiba lattices as well as chains, one might question the relevance of the results \new{ for Shiba chains in the previous section for experimentally relevant systems.}
%
%
In general, the phase diagrams of 1D and 2D systems are different --- there is no reason that a Shiba chain which is embedded into a 2D substrate behaves like the purely 1D system. Instead, one would expect a combination of a 1D phase diagram (as shown in panels (a) of Figs.\,\ref{fig:s000-chain}, \ref{fig:s100-chain}, \ref{fig:s000_s110-chain}) and the corresponding 2D phase diagram at $J=0$ (for extended $s$-wave see the first row of Fig.\,\,\ref{fig:ext_s-wave}), since the substrate is not covered by magnetic adatoms.

\begin{figure}[t!]
\centering
\includegraphics[width=0.98\columnwidth]{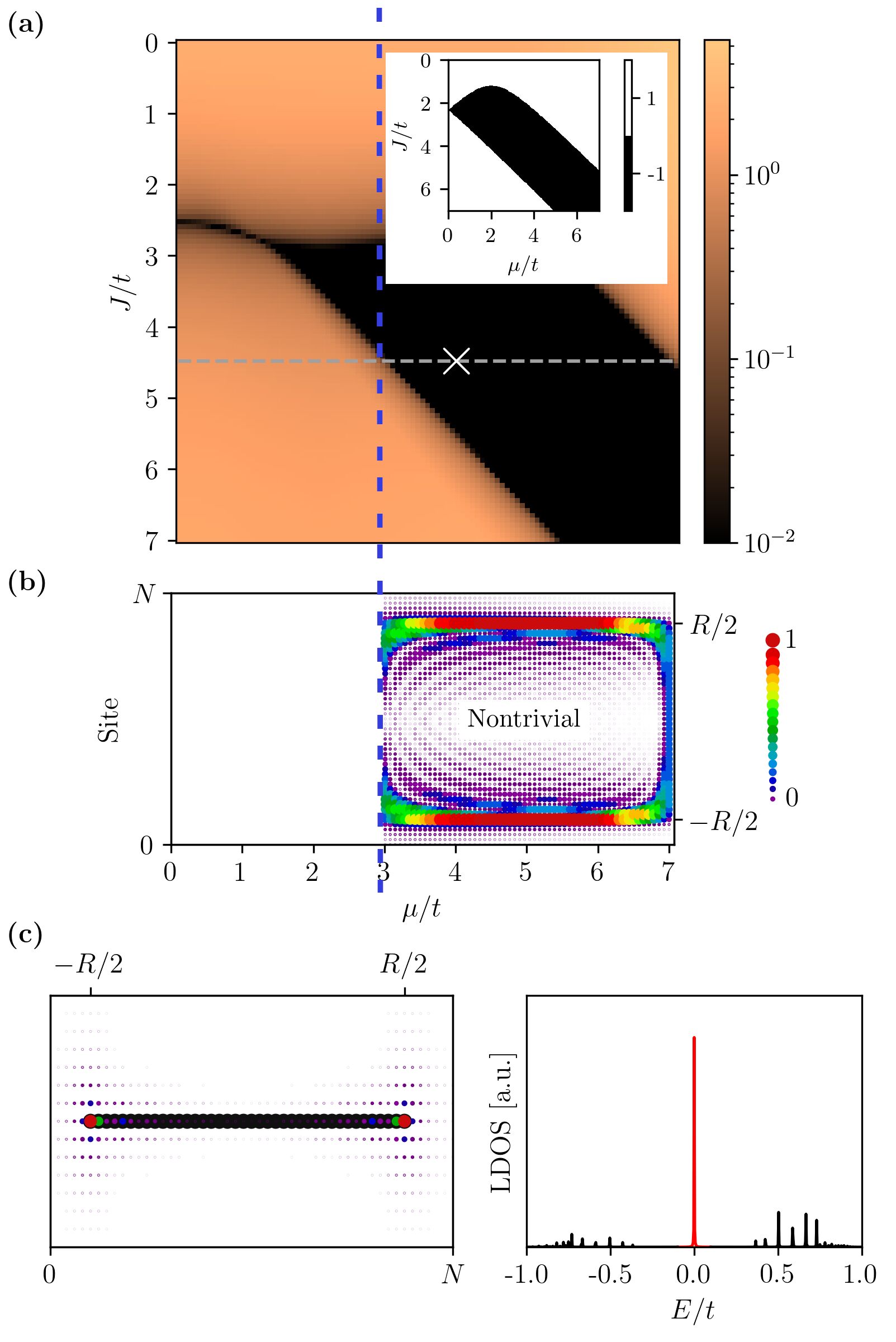}
\caption{
Shiba chain on a 2D substrate with conventional $s_{000}$ pairing. (a) Gap size as a function of $J/t$ and $\mu/t$. Note that gap closings can be either due to gapless bulk states of the chain or Majorana zero modes. Inset: Phase diagram of the purely 1D system for the same parameter ranges. (b) Eigenstates $|\psi|^2$  at $E=0$ which is integrated along the shorter $x$ direction (perpendicular to the chain) at $J=4.5t$. The plot reveals zero modes at the chain ends at $\pm R/2$, identifying the topological phase, see main text for details. 
\new{(c) Spatially resolved LDOS at $E=0$ (for $J=4.5t$, $\mu=4t$).
(d) Energy resolved LDOS at the end of the chain shown in panel (c); Majorana bound state is shown in red.}
Parameter used: $M\times N$ sites substrate with $M=15$, $N=51$, chain length of $R=40$, $\alpha=0.8$, $\Delta=1.2$.}
\label{fig:chain2D_s000}
\end{figure}

In the simplest case of an isotropic, nodeless $s$-wave substrate such a combined phase diagram is trivial: the $s_{000}$ substrate is always gapped at $J=0$. Thus the combined phase diagram should be identical to the purely 1D case. This is, however, not the case. In Fig.\,\ref{fig:chain2D_s000}\,(a) we show the real phase diagram of a chain of $R=40$ adatoms embedded into a larger superconducting substrate of size $N\times M$ with $N=51, M=15$. The topologically non-trivial region is shifted to larger $J$ and to significantly larger $\mu$. In panel (b), we also show the $E=0$ wavefunction $|\psi|^2$ (projected to a single line) as a function of $\mu$, in agreement with the phase diagram. \new{Panel (c) and (d) show the spatial LDOS at $E=0$ and the energy-resolved LDOS at the chain end, respectively, of a selected system (marked by the white $\times$ in the phase diagram (a)).}

\begin{figure}[t!]
\centering
\includegraphics[width=0.83\columnwidth]{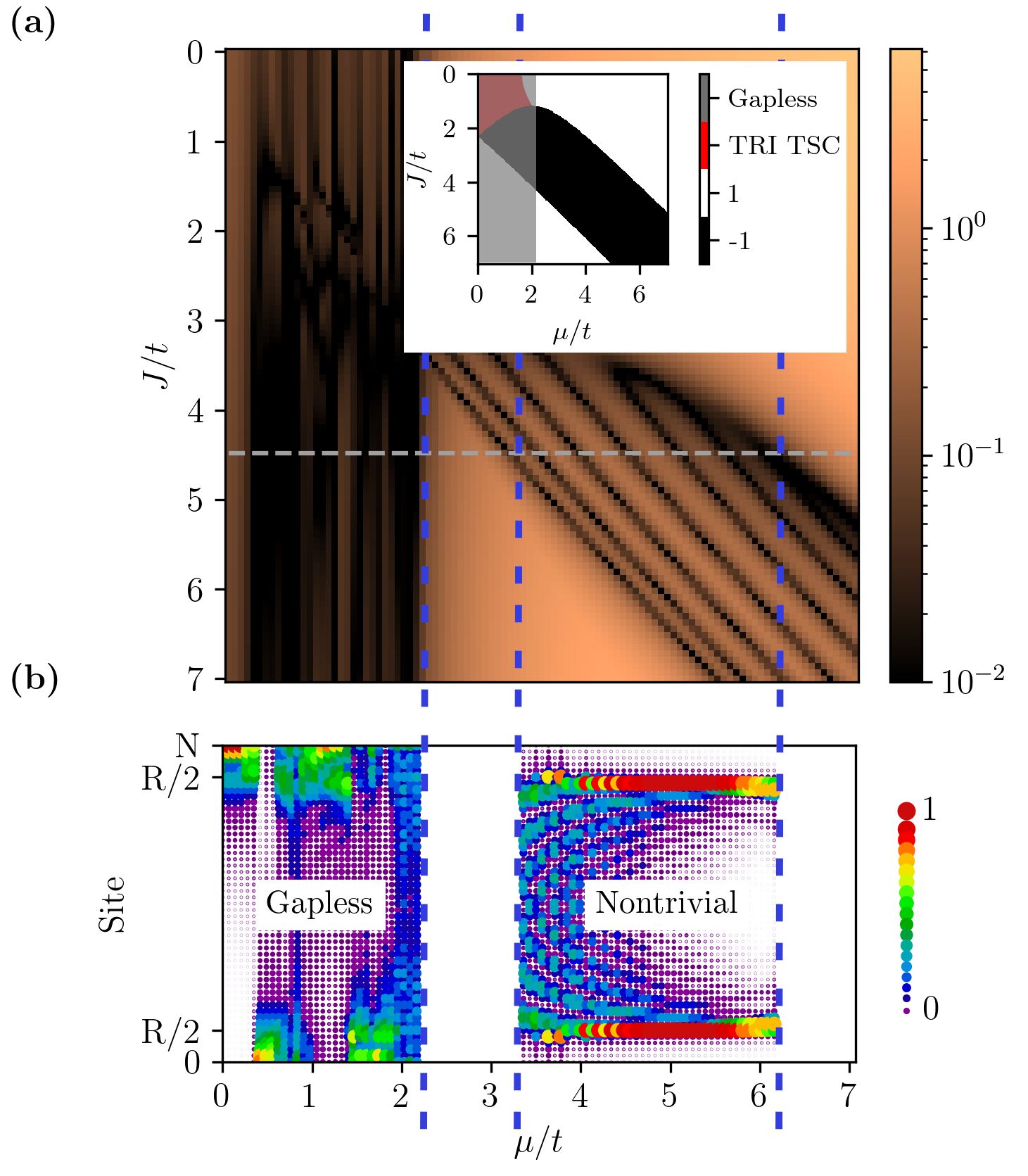}
\caption{
Shiba chain on a 2D substrate with extended $s_{100}$ pairing. (a) Gap size as a function of $J/t$ and $\mu/t$. Note that gap closings can be either due to the gapless substrate, gapless bulk states of the chain or Majorana zero modes. Inset: Combined phase diagram of the substrate with $J_{\rm substrate}=0$ and the purely 1D system for the same parameter ranges. (b) Eigenstates $|\psi|^2$ at or very close to $E=0$ which is integrated along the shorter $x$ direction (perpendicular to the chain) at $J=4.5t$. The plot reveals a gapless region (stemming from the substrate) and zero modes at the chain ends at $\pm R/2$, identifying the topological phase, see main text for details. Parameter used:  $M\times N$ sites substrate with $M=15$, $N=51$, chain length of $R=40$, $\alpha=0.8, \Delta=1.2$.}
\label{fig:chain2D_s100}
\end{figure}

As a second, more interesting example we focus on the extended $s_{100}$-wave substrate. Here the 2D phase diagram at $J=0$ is in a gapless regime for $\mu<2t$ and trivially gapped for $\mu>2t$. The superimposed 1D and 2D phase diagrams are
 schematically shown in the inset of Fig.\,\ref{fig:chain2D_s100}\,(a). As for the $s_{000}$ case, the real phase diagram (shown in Fig.\,\ref{fig:chain2D_s100}\,(a)) is different. While the gapless phase persists up to slightly larger values than $\mu=2t$, the topological phase is again shifted towards larger values of $J$ and $\mu$. There are no signs of the TRI topological phase anymore; the generic gaplessness of the substrate destroys the topological phase of the chain \new{for small values of $\mu$}. In panel (b), we show the wavefunction $|\psi|^2$ (projected to a single line) \new{at or very close to $E=0$} as a function of $\mu$ confirming the calculated phase diagram.

We also investigated the $s_{110}$ substrate (not shown here) with very similar results than those shown in Fig.\,\ref{fig:chain2D_s100}. The main difference is the more extended gapless region present up to $\mu=3.5t$, which is directly inherited from the 2D phase diagram Fig.\,\ref{fig:ext_s-wave}\,(b) at $J=0$.

Our results show that it is not sufficient to study the purely 1D case (at least not for realistic MSH structures). Not only have we observed a considerable change of parameters, also an entire phase (the TRI topological phase) has disappeared.
While we have considered here only a one-orbital toy model, we believe that our results are instructive and that they will apply to more realistic multi-orbital models as, for instance, obtained from {\it ab initio} methods. 

We mentioned before that in a truly 1D system the \(s_{100}\) and \(d\)-wave pairings are identical. For a chain embedded into a substrate that is of course no longer the case. We find, however, that for reasonable parameter ranges both $d$-wave substrates lead to a gapless phase diagram and no topological phases appear.

%
%
\subsection{Stability of TRI Topological Edge States}\label{topological-protection-of-tri-tsc-edge-states}

In both 1D and 2D, we have found a TRI topological phase in most MSH structures involving some type of extended $s$-wave superconducting substrate. Discovered by Zhang-Kane-Mele in 2013\,\cite{zhang_time-reversal-invariant_2013}, the authors pointed out that the edge states persist for finite Zeeman field $J$ until the bulk gap closes. Finite Zeeman field breaks, however, time-reversal symmetry, resulting in the loss of topological protection. Zhang-Kane-Mele concluded that the observed edge states are only protected as long as $J=0$, otherwise these edge modes are {\it unprotected}. \new{Zhang \ea\,\cite{zhang_time-reversal-invariant_2013} showed for the 1D case that the MZMs cannot be coupled by magnetic moments pointing in the $x$ or $z$ direction due to their spin structure. In contrast, magnetic moments along the $y$ direction couple them and immediately shift them to finite energy.
}

\begin{figure}[t!]
\centering
\includegraphics{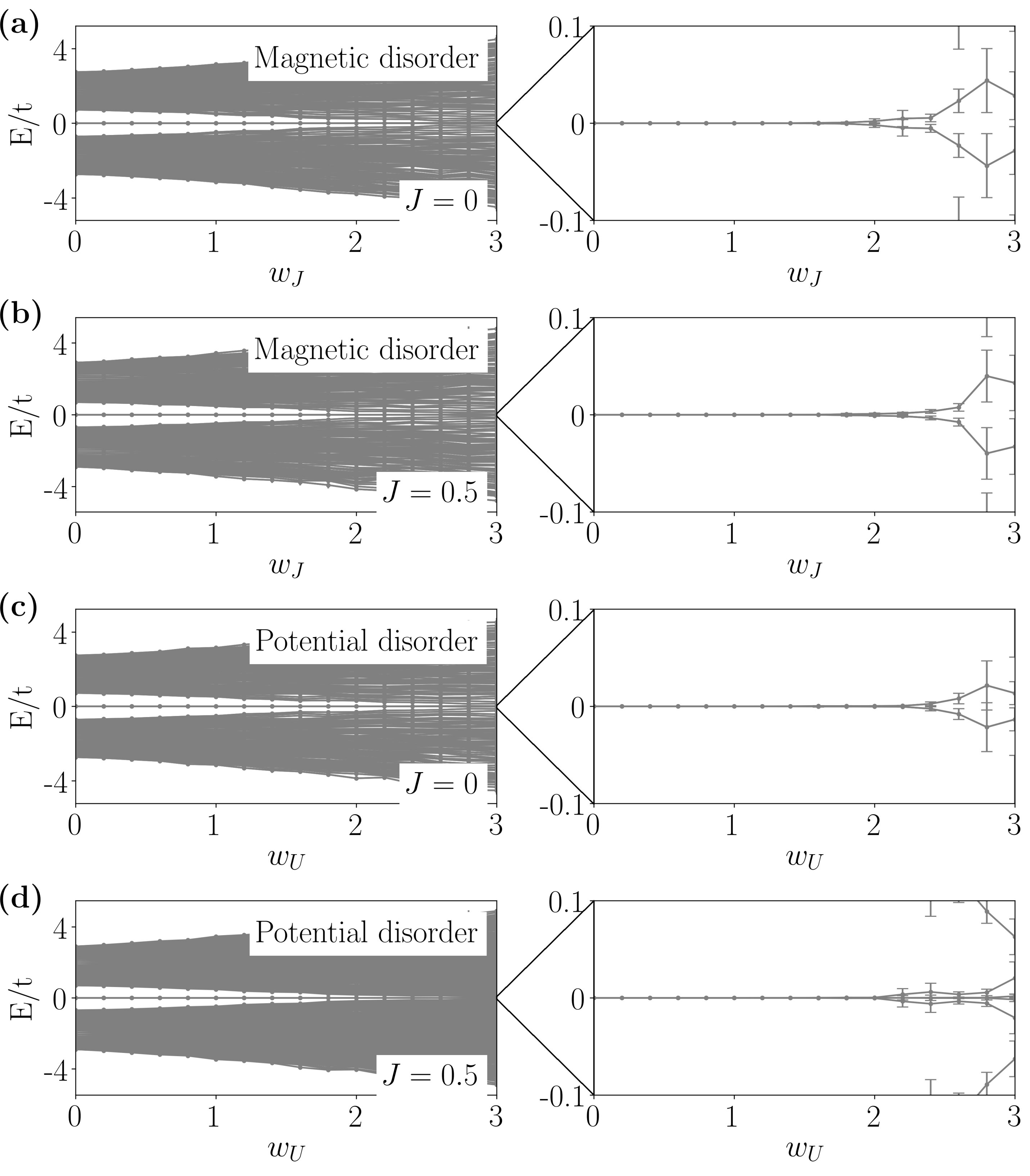}
\caption{Stability of topologically protected ($J=0$) and ``unprotected'' ($J=0.5$) helical Majorana bound states of the Shiba chain on $s_{100}$ substrate, purely 1D with OBCs imposed. (a,b) Random magnetic disorder with strength $w_J$. (c,d) Random potential disorder with strength $w_U$. Five disorder realizations, leading to the standard deviation shown for the ingap states.
Parameters used: chain length $R=50$, $\Delta=1.2t$, $\alpha=0.8t$, $\mu=0$.}
\label{fig:disorder1D}
\end{figure}

Here we inspect \new{these \emph{unprotected} end states} and probe their stability in the presence of random disorder, \new{assuming that the involved magnetic moments remain fixed in the $z$ direction}. We have shown that the TRI TSC phase is absent in the chain-in-2D geometry so consider just the purely 1D system with OBCs imposed. Since disorder effects are not the main focus of this paper we consider only one example with the $s_{100}$ substrate. We test the time-reversal protected state (\ie $J=0$) and compare it to the time-reversal broken (\ie $J=0.5$), but still ``TRI topological phase''. We compare random magnetic to random potential disorder, see Fig.\,\ref{fig:disorder1D}. For random potential disorder there are no significant differences between the $J=0$ and $J=0.5$ cases. While the bulk gap decreases continuously with increasing disorder strength, the Majorana bound states remain at $E=0$ up to a disorder strength $w_U \approx 2t$. For random magnetic disorder, we find very similar results. Again, up to a disorder strength $w_J \approx 2t$ the MZMs persist at zero energy for both $J=0$ and $J=0.5$.

%
%
\subsection{Anti-Shiba chains}\label{anti-shiba-chain}

\begin{figure}[t!]
\centering
\includegraphics{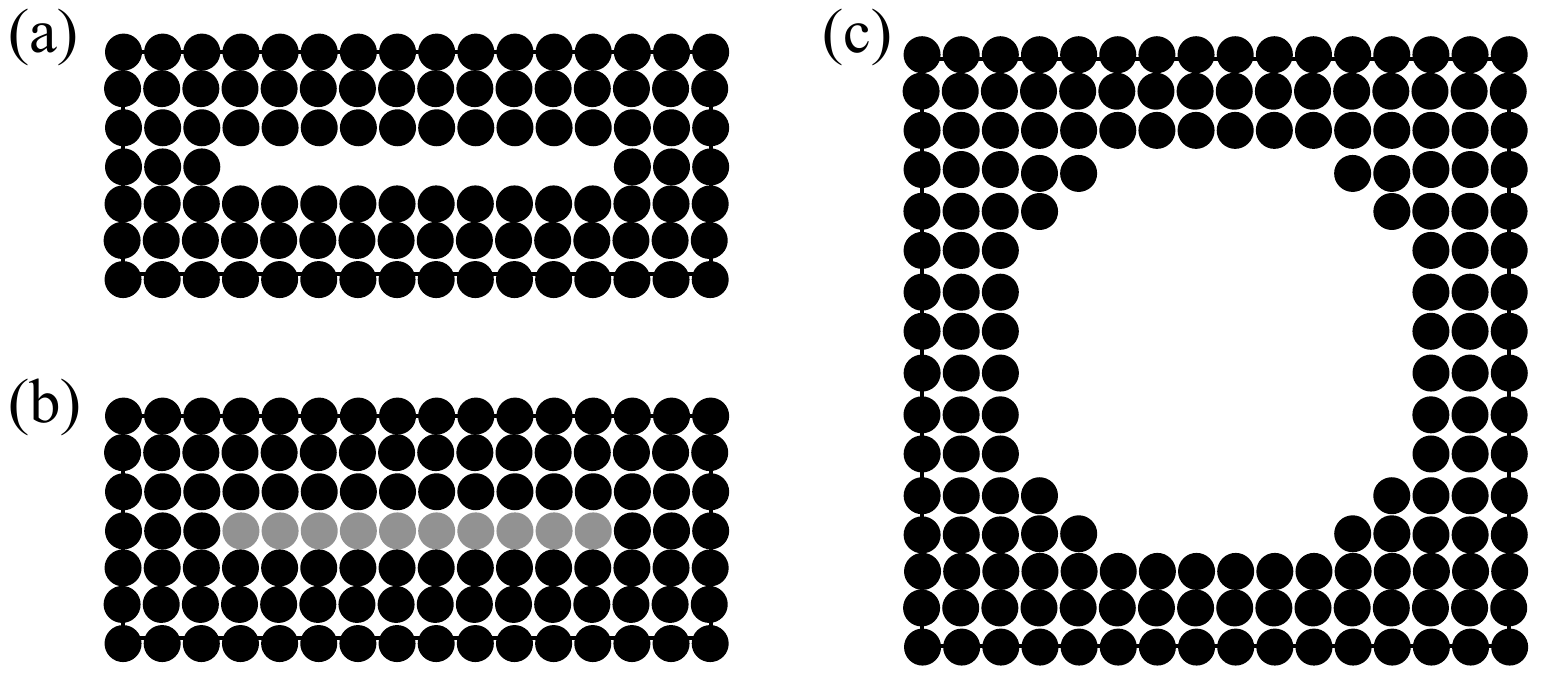}
\caption{(a) Anti-Shiba chain with $J_{\rm chain}=0$. (b) Variant of an anti-Shiba chain with $J_{\rm chain} \ll J_{\rm substrate}$. (c) Anti-Shiba island with $J_{\rm island}=0$.}
\label{fig:anti-shiba}
\end{figure}

In Sec.\,\ref{results1D}, we discussed the TRI topological superconducting state\,\cite{zhang_time-reversal-invariant_2013} with Kramer's pairs of localised Majorana zero modes at the chain ends, a variant of Majorana fermions which --- to our best knowledge --- has so far not been observed in experiments. Our previous analysis revealed, however, that for a realistic setup, \ie a chain embedded into a 2D substrate, the TRI topological phase is absent, see phase diagrams Figs.\,\ref{fig:chain2D_s000}\,(a) and \ref{fig:chain2D_s100}\,(a). In the following, we demonstrate that there is still a way to realize the TRI phase by building {\it anti-Shiba chains}. That is, one covers a larger two-dimensional region with magnetic adatoms (we consider areas of size $15\times 51$) except one row in the middle. We thus consider a ``missing'' chain of magnetic adatoms rather than a real chain, leading to the term {\it anti-Shiba chain}. Atom manipulation techniques allow to remove individual atoms or even an entire chain of surface atoms. The scenario under consideration is thus not unrealistic and experimentally feasible.

In what follows we focus on an extended $s$-wave substrate with $s_{100}$ pairing. We assume magnetic moments which are strong enough to fully gap the substrate. At the same time we assume $\mu=0$ (both for the substrate and the chain), thus the gapped substrate is topologically trivial. The chain, \ie the region where the magnetic adatoms have been removed, has $J=0$ and is in the TRI topological phase. Indeed we find four Majorana zero modes, two per chain end, showing that it is possible to stabilize the TRI TSC phase as an {\it anti-chain} setup. 

The substrate breaks time-reversal symmetry and the stability of the TRI topological phase with its Majorana edge modes is thus questionable. Due to the magnetic substrate, the topological protection is clearly lost and one would naively assume that each Kramer's pair of  Majorana zero modes gaps out in the presence of perturbations. Here we test the stability of the \new{Majorana bound states} in the presence of magnetic and potential random disorder as two likely sources of perturbations. First we restrict the disorder to the anti-chain, later we also consider disorder in the entire system, \ie both substrate and anti-chain. In addition, we also consider the situation that the chain is covered by magnetic atoms with much weaker coupling $J=0.5$ as a variation of the anti-chain with $J=0$. In principle, atom manipulation allows to place different sorts of atoms next to each other (such as Fe and Co or Mn), although our motivation to study $J=0.5$ is to rule out that the anti-chain with $J=0$ is a singular point rather than an extended phase.

\begin{figure}[b!]
\centering
\includegraphics[width=0.98\columnwidth]{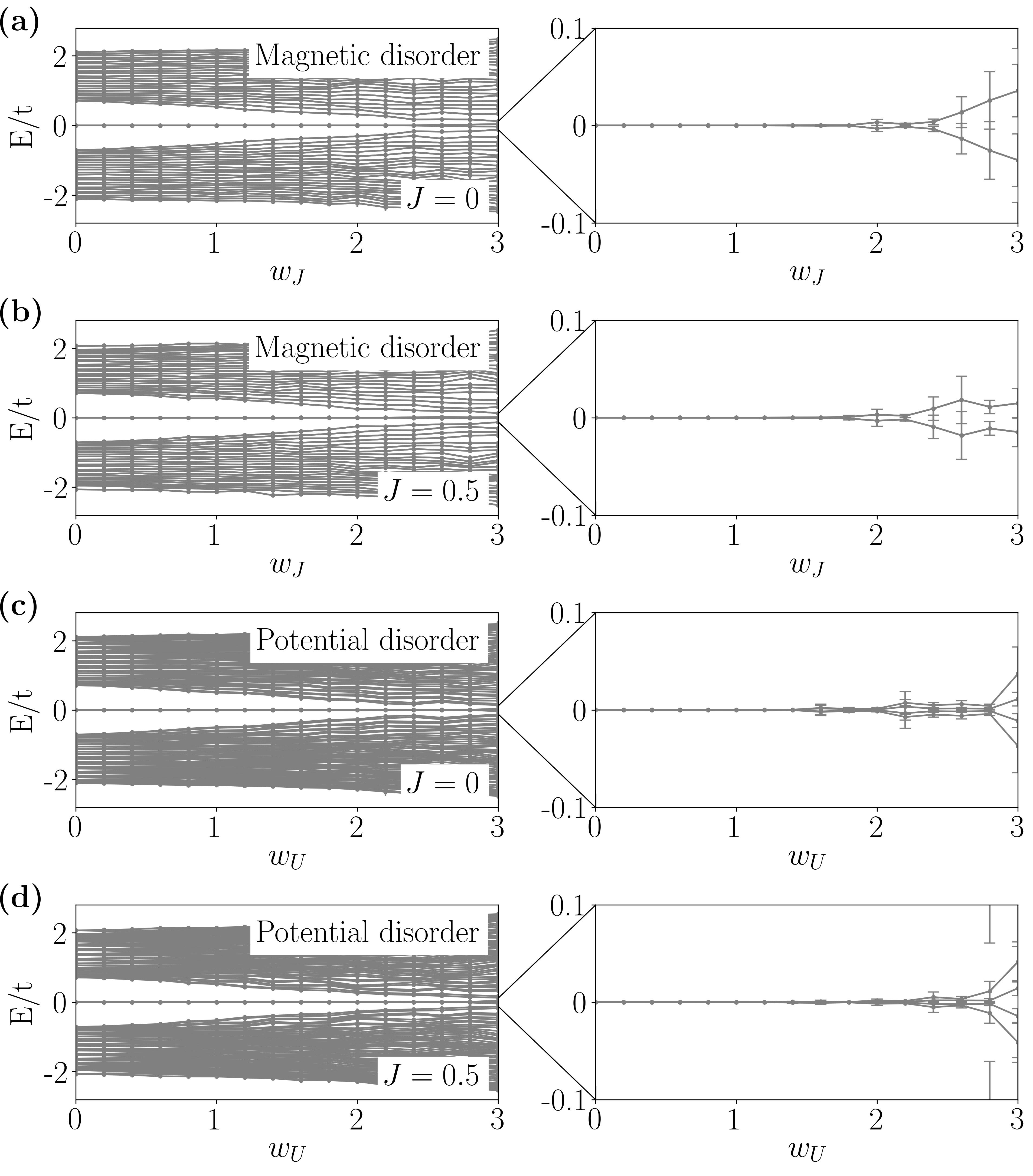}   
	\caption{Stability of ``unprotected'' helical Majorana bound states of the anti-Shiba chain [see Fig.\,\ref{fig:anti-shiba}\,(a)] in the presence of random magnetic (a,b) and potential disorder (c,d). Disorder is only placed on the chain, not on the surrounding substrate.
	``Unprotected'' TRI topological phase with $J=0$ (a+c) and   with $J=0.5$ (b+d). Five disorder realizations, leading to the standard deviation shown for the ingap states. Parameters used: $M\times N$ sites substrate with $M=15$, $N=51$, chain length of $R=40$, $\alpha=0.8$, $\Delta=1.2$.}
\label{fig:disorder-chain}
\end{figure}
\begin{figure}[t!]
\centering
\includegraphics[width=0.98\columnwidth]{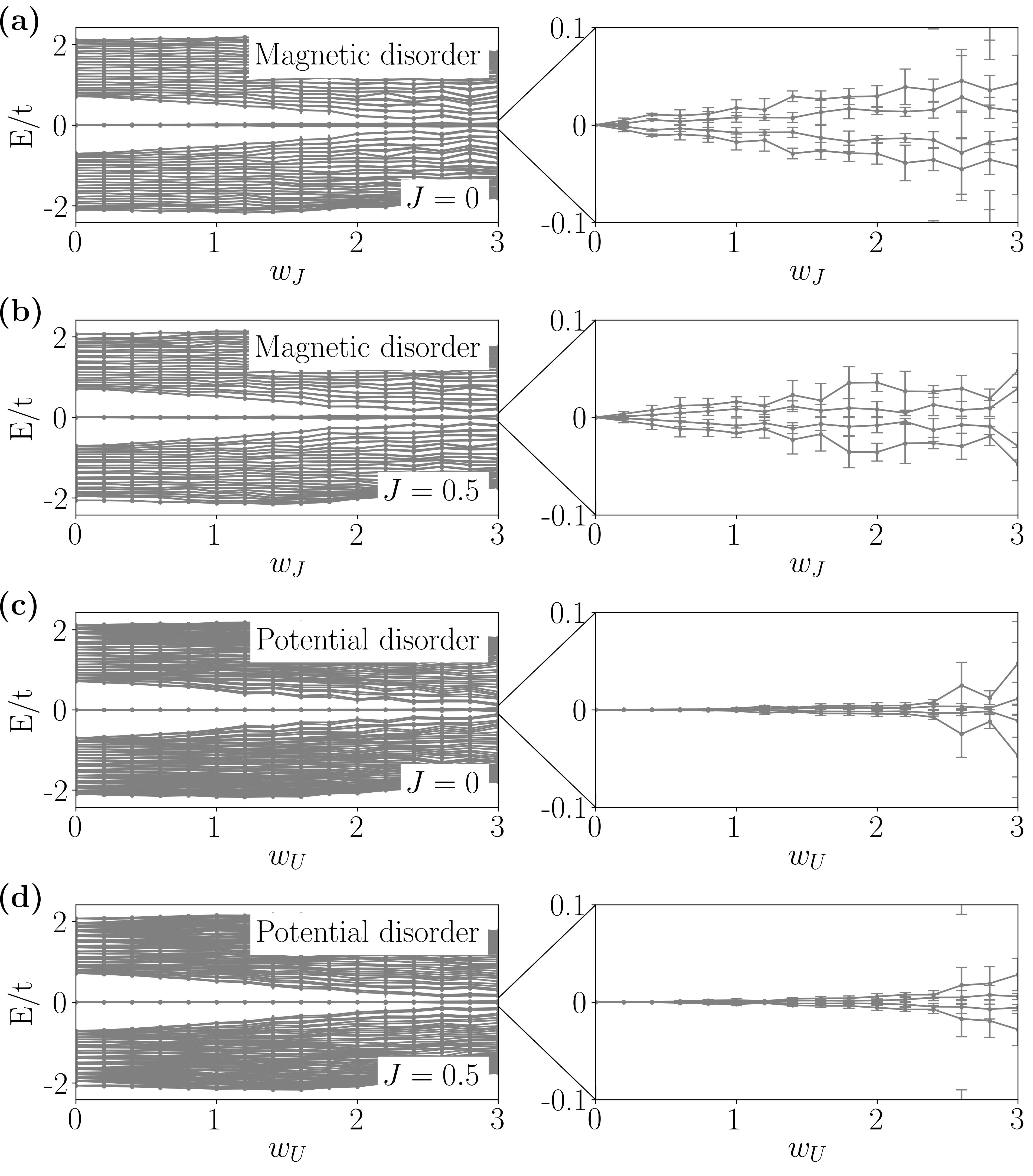}   
\caption{
Stability of ``unprotected'' helical Majorana bound states of the anti-Shiba chain in the presence of random magnetic (a,b) and potential disorder (c,d). Disorder is placed on the chain and the surrounding substrate.
 ``Unprotected'' TRI topological phase with $J=0$ (a+c) and   with $J=0.5$ (b+d). Five disorder realizations, leading to the standard deviation shown for the ingap states. Parameters used: $M\times N$ sites substrate with $M=15$, $N=51$, chain length of $R=40$, $\alpha=0.8$, $\Delta=1.2$.}
\label{fig:disorder-everywhere}
\end{figure}

We find that the  ``unprotected'' Majorana modes remain at zero-energy for both potential and magnetic disorder, when the disorder is restricted to the chain. This remains true for significant disorder strength: for magnetic random disorder $w_J<2t$ and for potential random disorder $w_U<2t$ the zero modes remain stable; at the same time we observe a continuous decrease of the bulk gap with increasing disorder strength. These results are summarized in Fig.\,\ref{fig:disorder-chain}.

Finally we investigate the case where the disorder is on the entire surface. For potential disorder, we still find the Majorana edge modes to persist at zero energy for disorder strengths as large as $w_U \approx t$. For magnetic random disorder, the Majorana modes immediately split to finite energies and are no longer stable. These results are summarized in Fig.\,\ref{fig:disorder-everywhere}. We note, however, that any topological phase protected by time-reversal symmetry is expected to break down in the presence of magnetic random disorder and the latter results are thus not unexpected at all. What is rather surprising, however, is the stability of the TRI topological phase in the presence of potential random disorder and also of magnetic disorder restricted to the chain area, since time-reversal symmetry is broken due to the substrate.

We note that the stability of TSC phases in normal MSH structures, \ie those which are characterized by a finite Chern number, was studied recently. In Ref.\,\onlinecite{mascot-19prb235102}, Shiba islands in the presence of potential and various types of magnetic random disorder were shown to be surprisingly robust.

%
%
\section{Discussion}\label{discussion}

Finding suitable MSH structures with large superconducting gaps is one of the main challenges on the way to unambiguously identify Majorana zero modes and eventually perform topological braiding. Here we have investigated several unconventional superconducting substrates with spin-singlet pairing, as realized in many of the known high-temperature superconductors. 

Standard $d$-wave superconductors such as cuprates are notoriously gapless due to their nodal lines of the $d_{x^2-y^2}$-wave order parameter; the (001) surfaces of this material class is thus not useful for MSH structures. Introducing strong
next-nearest-neighbour hopping opens a gap and a topologically nontrivial phase
becomes accessible at large chemical potential \(\mu\) and Zeeman
amplitude \(J\). These are Chern number \(\mathcal{C}=\pm 2\) phases. There
is currently no experimental evidence for any cuprate with such a pronounced second neighbor hopping.
Possibly strain or pressure engineering might lead to future experiments.

Candidate materials for chiral $d$-wave pairing might be water intercalated cobaltates\,\citep{takada_superconductivity_2003, PhysRevLett.91.097003} and twisted bilayer graphene as well as related heterostructures\,\cite{dai_twisted_2016, cao_unconventional_2018, shen_observation_2019}.
Also some of the iron-based superconductors were proposed to inherit chiral \(s+id\)-wave pairing\,\citep{chen_iron-based_2014, reid_d-wave_2012, kuroki_unconventional_2008}.

The most promising material family are the iron-based superconductors with their extended \(s\)-wave
pairing. Notable candidates are FeSe (\(T_c=8\)K, higher $T_c$ at high pressure)
\citep{hsu_superconductivity_2008, medvedev_electronic_2009, wang_evidence_2018},
FeSeTe (\(T_c=14.5K\))
\citep{wang_evidence_2018, zhang_observation_2018}, LaFeAsO
(\(T_c=26\)K)
\citep{kuroki_unconventional_2008, mazin_unconventional_2008} and
BaKFe\(_2\)As\(_2\) (\(T_c=38\)K)
\citep{li_nematic_2017, rotter_superconductivity_2008}. 
In addition, there are many
other iron-based superconductors with higher $T_c$; the pairing symmetry of these materials is, however, less clear and under debate. Examples include SrFFeAs (\(T_c=56\)K) \citep{wu_superconductivity_2009} and
monolayer FeSe (\(T_c=100\)K)
\citep{ge_superconductivity_2015, fan_plain_2015, ge_evidence_2019, jandke_unconventional_2019}.
There are likely many more candidate extended \(s\)-wave superconductors to be
found in iron-based materials. Outside of these, highly overdoped
monolayer CuO\(_2\) (\(T_c=91\)K) has also been proposed as extended
\(s\)-wave \citep{jiang_nodeless_2018, zhong_nodeless_2016}. 
Quite generally, larger transition temperatures $T_c$ usually correspond to larger superconducting gap sizes. For instance, FeSe on SrTiO$_3$(001) has a gap as large as 18meV\,\cite{ge_evidence_2019, jandke_unconventional_2019}. Gap sizes up to 20 meV have also been reported in LaFeAsO and LaFePO\,\citep{ishida_unusual_2008}.

The Shiba
island and chain MSH systems involving unconventional substrates presented here are very
promising and any of these materials might be candidate platforms for TSC heterostructures and Majorana physics.

\begin{figure}[t!]
\centering
\includegraphics{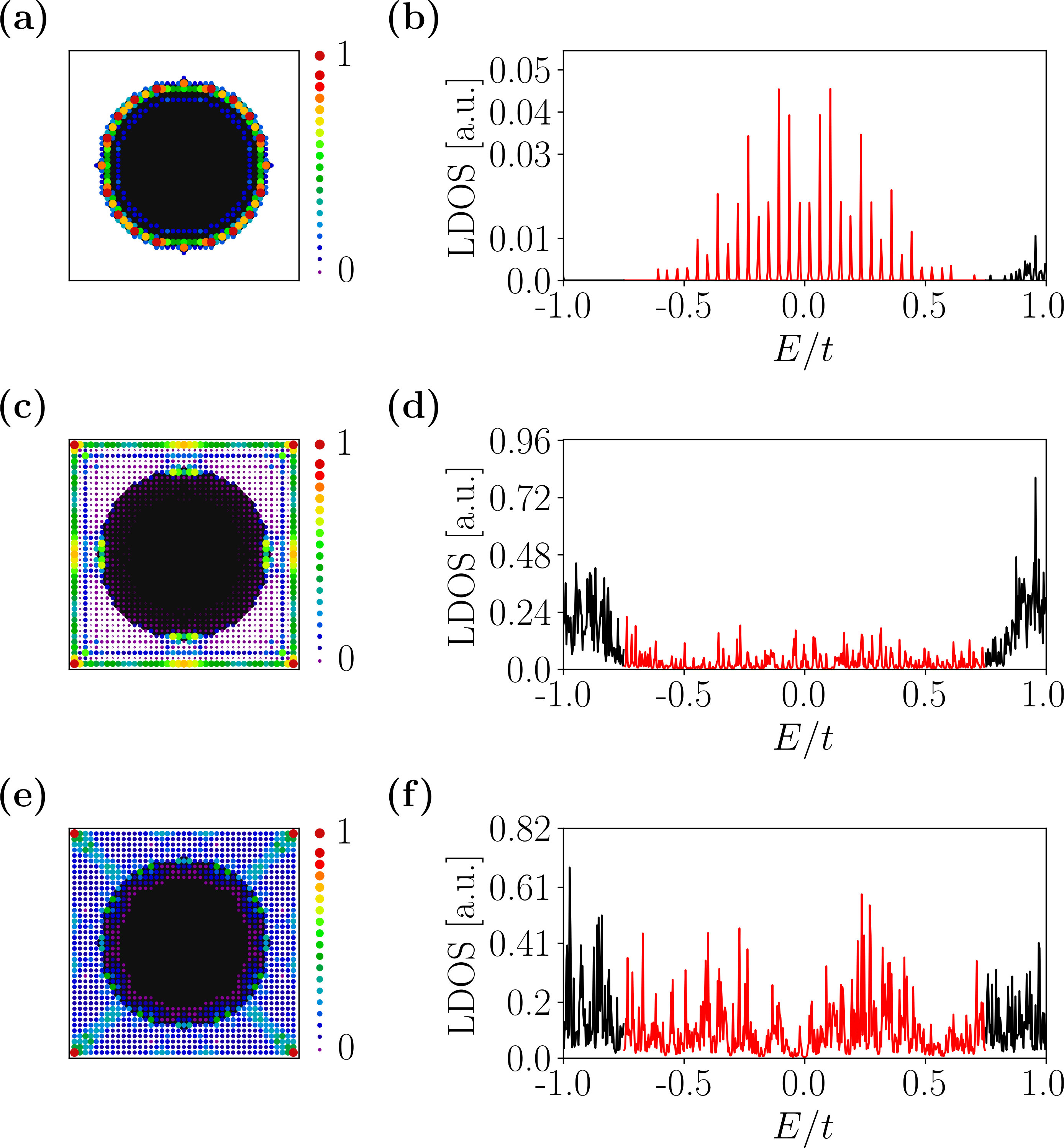}
\caption{
Examples for LDOS of Shiba islands with a magnetic substrate, a TRI topological substrate and, a gapless substrate. For all cases the substrate is $s_{000} + s_{100}$; the island is in the $\mathcal{C}=1$ phase; the geometry is a $41\times41$ lattice with an island of radius \new{$R=15$}; and parameters are $\alpha=0.8, \Delta=1.2, \gamma=0.001t$.
	(a,b) Gapped substrate leading to a perfect resolution of the chiral Majorana mode. Parameters used: $J_{\rm substrate} = 8t, J_{\rm island} = 5t, \mu = -2t$.
(c,d) Non-magnetic substrate i.e. gapless. The uncovered area realizing the TRI phase is too small to support a sufficiently large bulk gap, thus leading to a strong penetration of the edge mode between the topological phases. Parameters used: $J_{\rm substrate} = 0, J_{\rm island} = 5t, \mu = -2t$. (e,f) Gapless substrate with fully gapped Shiba island. The chiral Majorana mode fully hybridizes with the gapless substrate.  Parameters used: $J_{\rm substrate} = 0, J_{\rm island} = 5t, \mu = -0.25t$.}
\label{fig:other-substrates}
\end{figure}

In Sec.\,\ref{results2D} we showed energy-resolved and spatial LDOS plots for selected systems where a full gap leads to a reasonably well pronounced chiral edge mode. Actual experiments have \textit{a priori} a significant uncertainty about the material parameters. As a consequence, it is hard to predict where in the phase diagram the experiment will be located. In the following, we discuss three cases which are different from the scenarios considered in Sec.\,\ref{results2D}. For all examples we assume extended $s$-wave substrates; we note, however, that we do not expect qualitative differences for other type of substrates with the same properties (\ie magnetic, TRI topological or gapless substrates). In Fig.\,\ref{fig:other-substrates}\,(a) we consider an MSH structure where the entire substrate is covered with magnetic adatoms, and the Shiba island is defined by a different (weaker) type of magnetic adatom. This is a variant of the anti-Shiba island sketched in Fig.\,\ref{fig:anti-shiba}\,(c). Due to the magnetic moments on the substrate, there is a large bulk gap. The different magnetic coupling strength on the island realizes a topologically non-trivial phase and the concomitant chiral edge modes are strongly localized with a very short penetration length [see Fig.\,\ref{fig:other-substrates}\,(a)]. As a consequence, also the energy-resolved LDOS peaks of the edge modes \new{are well-pronounced and feature equal energy spacing known from a linear $E(k)$ dependence [see Fig.\,\ref{fig:other-substrates}\,(b)].}

Next we consider a Shiba island where the substrate is not in a gapped, topologically trivial phase but instead in the TRI topological phase. An educated guess would tell us that between two gapped, distinct topological phases we will see edge modes. Since we consider the same setup as in all previous examples, the uncovered substrate is rather small compared to the island, causing a rather large penetration length of the topological edge modes. As a result, we observe an almost gapless substrate and the edge mode(s) is (are) no longer well defined [see Fig.\,\ref{fig:other-substrates}\,(c)]. Correspondingly, the energy-resolved LDOS is not different from bulk LDOS anymore [see Fig.\,\ref{fig:other-substrates}\,(d)]. The third example we consider is a gapless substrate. For a Shiba island in the topological phase ($C=1$), the corresponding chiral edge modes completely hybridize and delocalize with the gapless states of the substrate. The spatial LDOS plot does not show any signs of the edge mode anymore [see Fig.\,\ref{fig:other-substrates}\,(e)].
We note that this scenario corresponds to what had been found for Shiba islands in Fe/Re(0001) [see Supplement of Ref.\,\onlinecite{palacio-morales_atomic-scale_2019})].
 As a consequence, also the energy-resolved LDOS is indistinguishable from a bulk LDOS. \new{Note that the red region in Fig.\,\ref{fig:other-substrates}\,(d) and (f) corresponds to the local gap size in the center of the island.}
 
\new{These results demonstrate that,} despite the many topological phases shown in the various phase diagrams presented in this paper, it remains a fine-tuning issue of the relevant parameters to find chiral Majorana modes in experiments with a convincing spectral resolution\,\cite{palacio-morales_atomic-scale_2019}.

%
%
\hypertarget{conclusion}{%
\section{Conclusion}\label{conclusion}}

Magnet-superconductor hybrid (MSH) systems are a leading platform for
engineering topological superconductors and Majorana fermions. These systems are thus a promising
approach to fault-tolerant quantum computing due to the non-Abelian
exchange statistics of the Majorana zero-modes found at system
boundaries. There is good experimental evidence for Majorana fermions in MSH structures with superconducting lead and rhenium substrates. However, some of these systems suffer from very small
superconducting gap sizes, spoiling the spectral resolution in experiments.
We have investigated various unconventional superconducting substrates with spin-singlet
pairing (including $d$-wave, chiral $s$ and $d$-wave as well as extended $s$-wave) with potentially much higher transition temperatures and larger gap sizes. Essentially any type of unconventional substrate can be used to realize Shiba chains or islands, as long as a full superconducting gap can be guaranteed.
In particular the iron pnictides and  chalcogenides are a promising family of candidate materials. 
We have also found time-reversal invariant topological superconductivity of the Zhang-Kane-Mele type which is realized in the extended $s$-wave MSH structures. Our results suggest that even in the presence of magnetic moments breaking time-reversal explicitly, the helical Majorana modes can be robust in the presence of random disorder, although they are not topologically protected anymore. We have also shown that results obtained for purely one-dimensional geometries fail to describe the experimental Shiba chain systems, where the chain of magnetic adatoms is embedded into the two-dimensional substrate. We have calculated correct phase diagrams for these cases.

\begin{acknowledgements}
The authors acknowledge interesting discussions with J.\ Wiebe, H.\ Kim, L.\ Schneider, and R.\ Wiesendanger.
SR acknowledges support from the ARC through DP200101118.
\new{EM and DKM acknowledge support from the US Department of Energy, Office of Science, Basic Energy Sciences, under award no.\ DE-FG02-05ER46225.}
\end{acknowledgements}

\vspace{4cm}


\appendix*

\hypertarget{extended-s-wave-pairing-notation}{%
\section{\texorpdfstring{Extended \(s\)-wave pairing
notation}{Extended s-wave pairing notation}}\label{extended-s-wave-pairing-notation}}

Wenger \ea introduced a notation for describing the order
parameters of unconventional superconductors \citep{wenger_d_1993}. This
follows from the characters of \(D_{4h}\), and considering the general
spin rotationally invariant Hamiltonian expanded to linear order. The
extended \(s\)-wave expansion is
\begin{widetext}
\begin{align}
\Delta_k^{s^+} = \sum_{r_1=0}^\infty \sum_{r_2=0}^{r_1} \sum_{r_3=0}^\infty a_{r_1 r_2 r_3}^{s^+} (\cos k_x r_1 \cos k_y r_2 + \cos k_x r_2 \cos k_y r_1) \cos k_z r_3\ .
\end{align}
\end{widetext}
We are only interested in 2D and set $k_z=0$ corresponding to $r_3=0$.
For the lowest order ($r_1=r_2=0$) we get $\Delta_k^{s^+} = a_{000}^{s^+} = \Delta_{000}$, \ie plain $s$-wave. Similarly, the next order $r_1=1, r_2=0$ yields $\Delta_k^{s^+} = a_{100}^{s^+} (\cos k_1 + \cos k_2) = \Delta_{100} (\cos k_x + \cos k_y)$. The indices of $s_{abc}$ and $\Delta_{abc}$ then correspond to the order factors $r_{1,2,3}$ and so to the different orders of the expansion.

\bibliography{highTmajos.bib}

\end{document}